\PassOptionsToPackage{dvipsnames}{xcolor}
\documentclass[a4paper,11pt]{article}
\DeclareUnicodeCharacter{2212}{\textminus}
\usepackage{jcappub}
\usepackage{float}

\usepackage{orcidlink}
\usepackage{amsmath}
\usepackage{booktabs}
\usepackage{tabularx}
\usepackage{makecell}

\usepackage{bm}

\title{\texttt{GAME}: Genetic Algorithms with Marginalised Ensembles for model-independent reconstruction of cosmological quantities}

\author[a,b,c]{M. Peronaci\orcidlink{0009-0009-9719-0173},}
\author[d,e]{M. Martinelli\orcidlink{0000-0002-6943-7732},}
\author[f]{S. Nesseris\orcidlink{0000-0002-0567-0324}}
\affiliation[a]{Dipartimento di Fisica, Università di Roma ``Tor Vergata'', Via della Ricerca Scientifica 1, 00133, Roma, Italy}
\affiliation[b]{Istituto Nazionale di Fisica Nucleare (INFN), Sezione di Roma 2, Via della Ricerca Scientifica 1, 00133, Roma, Italy}
\affiliation[c]{Dipartimento di Fisica, Università di Roma ``Sapienza'', Piazzale Aldo Moro 5, 00185, Roma, Italy}
\affiliation[d]{INAF - Osservatorio Astronomico di Roma, via Frascati 33, 00078 Monte Porzio Catone, Italy}
\affiliation[e]{Istituto Nazionale di Fisica Nucleare INFN, Sezione di Roma, P.le A. Moro 5, I-00185, Roma, Italy}
\affiliation[f]{Instituto de F\'isica Te\'orica (IFT) UAM-CSIC, Calle Nicol\'as Cabrera 13-15, Campus de Cantoblanco UAM, 28049 Madrid, Spain}

\emailAdd{matteo.peronaci@roma2.infn.it}

\abstract{\enlargethispage{2\baselineskip} Genetic Algorithms (GA) are a versatile tool for stochastic optimisation and non-parametric symbolic regression, already widely used in cosmology. They are capable of reconstructing analytical functions directly from data points without introducing new physical models. A limitation of this approach is that while the reconstructed function is very efficient at reproducing the behaviour of the data points, non-observable quantities involving derivatives are particularly sensitive to stochasticity, hyperparameters, and the GA's chosen best-fit function, risking the algorithm getting stuck in a local minimum. In this work we propose an update to the GA methodology for the reconstruction of analytical functions that involves computing a weighted average of an ensemble of GA configurations (\texttt{GAME}). We define the weights via a quantity that accounts for both the goodness-of-fit of the points and the smoothness of the resulting function. We also present a practical method to analytically estimate and correct the errors on the averaged function by combining a path-integral approach with an ensemble variance. We demonstrate the improvement offered by \texttt{GAME} methodology on a generic test function. We then apply the new methodology to a non-parametric reconstruction of the Hubble rate $H(z)$ using Cosmic Chronometers data and, assuming a flat Friedmann-Lemaître-Robertson-Walker background and General Relativity, we infer the corresponding dark energy equation of state $w(z)$. As an additional test, we apply the updated methodology to reconstruct the distance modulus $\mu(z)$ from Type Ia Supernovae data and its corresponding derived $H(z)$ and $w(z)$. Finally, we show that current data produce results compatible with $\Lambda$CDM, and that future cosmological surveys will allow GA reinforced with \texttt{GAME} methodology to become an even more competitive tool for discriminating between different models.}

\begin{document}
\maketitle

\flushbottom

\section{Introduction}
\label{sec:intro}
The standard cosmological model, $\Lambda$CDM, has achieved unparalleled success in describing a wide range of observations \citep{Peebles:2003}, from the anisotropies of the Cosmic Microwave Background (CMB) to the large-scale distribution of galaxies. Within Einstein's equations of General Relativity (GR), this success relies on the introduction of Cold Dark Matter (CDM) and dark energy in the form of a cosmological constant $\Lambda$. Despite its effectiveness, the model still faces major open questions: the unknown nature of the CDM component and of $\Lambda$, as well as prominent observational tensions in the measurements of the Hubble constant $H_0$ \citep{Perivolaropoulos:2022, CosmoVerseNetwork:2025alb}. While these issues suggest that $\Lambda$CDM might not be a complete description of the universe, there is currently no compelling alternative model that satisfactorily solves this tension \cite{Schoneberg:2021}. Recent analyses from DESI \cite{DESI:2024} have revived the discussion on a possibly evolving dark energy equation of state, with combinations of Baryonic Acoustic Oscillations (BAO), CMB, and Type Ia Supernovae (SNe Ia) data showing a mild preference for $w(z)\ne-1$. Taken together, these hints suggest that, rather than committing a priori to a specific parametric extension, it is worth letting the data themselves guide the search for departures from $\Lambda$CDM.

This necessitates the search for robust model-independent and non-parametric reconstructions that can probe cosmological functions directly from data, minimising theoretical priors and parametrisations. A particularly promising direction is to test the consistency between the cosmic expansion history and the properties of dark energy inferred from it. In GR, the background expansion and the physical content of the universe are tightly linked: if we reconstruct the expansion rate $H(z)$ and infer an effective dark energy equation of state $w(z)$, any significant and statistically robust deviation from $w(z)=-1$ could point to new physics. The difficulty lies in the fact that consistency tests based on non-observable quantities often require derivatives of reconstructed functions, making them especially sensitive to hyperparameters, overfitting, and the smoothness of the best-fit function. In fact, even when two reconstructions of $H(z)$ are statistically indistinguishable on data, their derivatives can differ enough to alter the conclusions on $w(z)$. Extracting reconstructions that are not only good fits to the data, but also produce stable derivatives, is therefore one of the central challenges of model-independent cosmology in the precision era. It is precisely this challenge that the present work addresses.

Various non-parametric methods have been employed to address this, most notably Gaussian Processes (GP) \citep{Seikel:2012, Seikel:2013}. A GP provides a fully bayesian reconstruction of a function and its derivatives, requiring the definition of a kernel function, which is in charge of controlling the correlations between different points of the reconstructed function \cite{Favale:2024dain}. However, although GP enable model-independent reconstructions by choosing a zero mean function to minimise cosmological assumptions, they can be sensitive to the choice of the kernel hyperparameters \citep{Perenon:2022}. In this work, we focus on Genetic Algorithms \citep{Bogdanos:2009, Nesseris:2012}, a symbolic-regression technique capable of reconstructing analytical functions from data without introducing new physical parameters. Unlike GP, GA return explicit functional forms that can be more easily interpreted than kernel-based representations. 

However, a known limitation is that the GA best-fit function can reproduce the data extremely well, yet non-observable quantities involving derivatives can be fragile. Indeed, different GA runs (or different hyperparameter configurations) may generate similarly good fits that present noticeably different derivatives. This methodology has been successfully applied across a range of cosmological contexts, including among the others null tests of the cosmological constant model \cite{Nesseris:2022Euclid, Arjona:2021, Nesseris:2010}, the construction of emulators for complex observables \cite{Aizpuru:2021vhd, Orjuela-Quintana:2022nnq, Orjuela-Quintana:2023uqb, Orjuela-Quintana:2024hha, Bartlett:2024}, performing searches for new physics beyond $\Lambda$CDM \cite{Akrami:2009hp,Arjona:2024cex,Arjona:2024dsr,Arjona:2020kco}, and the design of quantum optics experiments and related quantum optimization algorithms \cite{Gangopadhyay:2023nli, Acampora:2023pti}. 

Other applications of the GA in cosmology include using it to optimise Neural Networks \cite{gomez2023neural}, perform cosmological parameter estimation \cite{medel2023cosmological}, probe for cosmological tensions \cite{gangopadhyay2023phantom}, to analyse the data from precision measurements of charged cosmic rays \cite{luo2020genetic}, explore the string landscape \cite{bao2017fast}, optimise ionospheric models \cite{li2020advanced}, see also Ref.~\cite{charbonneau1995genetic} for a review of applications in astronomy and astrophysics. Finally, in a similar vein, symbolic regression has long been used in cosmology and particle physics to explore beyond the standard model physics \cite{abdussalam2025symbolic}, perform automated discovery of physical laws \cite{udrescu2020ai,bartlett2023exhaustive,tenachi2023deep}, automatically classify astronomical objects \cite{llorella2025exploring},  include numerical parameter priors in such a way that models may be fairly compared though the Bayesian evidence \cite{bartlett2023priors}, see also Ref.~\cite{angelis2023artificial} for a broader review.

The reason why derivatives are the natural bottleneck of symbolic regression reconstructions is structural: the fitness of any candidate function is evaluated only on the observable, $y_i$, sampled at a finite set of points $x_i$. Local noise may induce spurious local structure in the reconstructed function and can lead to unstable derivatives. Two functions that differ by a high-frequency oscillation of small amplitude, small enough not to be penalised by the $\chi^2$ given the data errors, can have very different derivatives, with the difference amplified at the boundaries of the data range. The GA best-fit reconstructions therefore may inherit a derivative that is essentially a coin toss between similar candidates. The central methodological contribution of this work is to suppress this fragility. We therefore propose an update to the baseline GA\footnote{\url{https://github.com/snesseris/Genetic-Algorithms}} approach, which considers just a single global minimum best-fit function, based on a model-averaging procedure over an ensemble of GA reconstructions. The key idea is to compute a weighted average of many GA best-fit functions in order to reduce the uncertainty due to stochasticity or to a particular choice of hyperparameters, defining weights through a combined estimator that accounts not only for the goodness-of-fit (via $\chi^2$) but also for smoothness (via a roughness penalty). We refer to this marginalised ensemble methodology as \texttt{GAME}\footnote{\url{https://github.com/matteoperonaci/GAME}} (Genetic Algorithms with Marginalised Ensembles).

In addition, since a model-averaged reconstruction requires a consistent uncertainty estimate, we introduce a practical method to estimate and combine both the measurement-driven statistical uncertainty, computed through a path-integral uncertainty approach \citep{Nesseris:2012}, and an additional configuration-driven uncertainty quantified by the weighted variance across the ensemble. We validate the improvement on a generic test function $f(x)$ and then apply \texttt{GAME} to the reconstruction of the Hubble rate $H(z)$ using Cosmic Chronometers (CC) data and to the reconstruction of the distance modulus $\mu(z)$ using SNe Ia. Finally, assuming a Friedmann-Lemaître-Robertson-Walker (FLRW) background and GR, we derive the corresponding dark energy equation of state $w(z)$ (from CC), the Hubble rate $H(z)$ (from SNe Ia), and compare them with the $\Lambda$CDM expectation. To further validate the \texttt{GAME} methodology and assess its robustness, we test it using mock data under control data-quality conditions.

This paper is structured as follows. In Section~\ref{sec:GAmethodology} we review the baseline GA framework and present the \texttt{GAME} marginalisation methodology, then describe the corresponding total uncertainty estimation. In Section~\ref{sec:CosmoApp} we apply \texttt{GAME} to cosmological observables, reconstructing $H(z)$ from CC data and deriving the dark energy equation of state $w(z)$, $\mu(z)$ from SNe Ia data and deriving the Hubble rate $H(z)$, and we then test performance on mock datasets. We conclude in Section~\ref{sec:conclusions} with a summary of the results and possible extensions of the work.

\section{Genetic Algorithms methodology}\label{sec:GAmethodology}
Genetic Algorithms (GA) are a model-independent machine learning tool for symbolic regression. They follow the principles of biological evolution via natural selection, where a population of mathematical functions, often referred to as chromosomes, evolves over time. In this context, a chromosome represents the analytical string of a function, and this evolution occurs under the influence of operations such as crossover (the combination of two different chromosomes to create an offspring) and mutation (a random change in one chromosome) \citep{Nesseris:2012}.

Although in computer science the evolution of analytical expressions through evolutionary operators is formally known as \textit{Genetic Programming}, we use the term ``GA'' throughout this work to remain consistent with the cosmological literature in which these methods were first introduced. Whenever we mention a ``baseline GA'' in this paper, we specifically refer to the standard cosmological implementation of symbolic regression as described in \cite{Bogdanos:2009, Nesseris:2012}.

The algorithm begins with an initial population of candidate individuals: $N_{\rm pop}$ functions randomly generated (analogous to traditional Monte Carlo approaches) based on a defined set of basic functions (sin, cos, log, polynomial, exp, etc.) and operators ($+,\, -,\, \times,\, \div$). Together, these elements constitute a set we refer to as the \texttt{grammar}. The algorithm is fed with $N_{\rm data}$ points in the form ($x_i,\, y_i,\, \sigma_i)$ and establishes the adequacy of each function to represent the data points by calculating the $\chi^2$ which, for uncorrelated measurements, is:
\begin{equation}\label{eq:chi2_uncorr}
    \chi^2 = \sum_{i=1}^{N_{\rm data}}\left(\dfrac{f_j(x_i)-y_i}{\sigma_i}\right)^2,\qquad \text{with}\quad j=1,\dots,N_{\rm pop},
\end{equation}
where $f_j(x_i)$ is one of the $N_{\rm pop}$ functions generated by the GA. The $\chi^2$ is then computed for each individual function in the population.

After generating the first population of functions, the algorithm starts producing the next generation by selecting a percentage of the best individuals (i.e., the functions with the minimum $\chi^2$ values) from the previous population. Once selected, these optimal individuals are combined through crossover and mutation to produce offspring. After the new population is generated, the algorithm repeats until a termination goal is achieved, such as a minimum $\chi^2$ value or a maximum number of generations $N_{\rm gen}$. Finally, the best-fit analytical function returned at the end of the iteration is the one presenting the minimum $\chi^2$.

The grammar, number of individuals in the population, selection percentage, crossover and mutation probability, number of generations, and stochastic seed constitute the hyperparameters of the GA; they affect how fast the GA converges to the best-fit and the complexity of the final expression.

\subsection{Practical example} 
We present a simplified example of a baseline GA iteration to explain how it determines the best-fit function given a dataset of $N_{\rm data}$. The following hyperparameters are fixed:
\begin{itemize}
  \setlength{\itemsep}{4pt}
  \setlength{\parskip}{0pt}
  \setlength{\parsep}{0pt}
  \setlength{\topsep}{0pt}
    \item $N_{\rm gen} = 100$;
    \item $N_{\rm pop} = 3$;
    \item \texttt{grammar} = polynomial, exponential, trigonometric;
    \item \texttt{selection rate} $= 2/3$;
    \item \texttt{crossover rate} $= 0.8$;
    \item \texttt{mutation rate}  $= 0.3$;
    \item \texttt{random seed} $=12345$;
\end{itemize}
The first population $\mathcal{F}(n_{\rm gen}=0)$ (with $n_{\rm gen} = 0,\ldots, N_{\rm gen}$) of $N_{\rm pop}$ functions is generated randomly, according to the \texttt{random seed} and the \texttt{grammar}. Let us assume the initial $N_{\rm pop}$ functions are:
\begin{align}
    \mathcal{F}(0) = \left\{x^2, \exp{x}, \sin{x}\right\}.
\end{align}
Initially, the algorithm computes the $\chi^2$ for each function in $\mathcal{F}(0)$ using Equation~\eqref{eq:chi2_uncorr} with the data points. Let us assume we obtain:
\begin{align}
    \boldsymbol{\vec{\chi}^2}(0) = \left\{300, 200, 600\right\}.
\end{align}
The algorithm sorts the population with respect to $\chi^2$. Since the fixed \texttt{selection rate} is $2/3$, it selects the two best functions out of the three. These are, respectively, the second ($\exp{x}$) and the first ($x^2$). Once the best functions are sorted and selected, the first generation begins, and the algorithm combines and modifies them via crossover and mutation, replacing all non-selected function slots (in this case, just the third remaining slot). One might obtain:
\begin{align}
    \mathcal{F}(1) = \left\{\exp{x}, x^2, \dfrac{\exp{x^3}}{x^2}\right\}.
\end{align}
Once $\mathcal{F}(1)$ is generated, the algorithm computes $\boldsymbol{\vec{\chi}^2}(1)$ and repeats the process. Finally, at the final generation (in our case $N_{\rm gen}=100$), we will have a population $\mathcal{F}(n_{\rm gen}=100)$ consisting of analytical and likely complex functions, alongside the final vector $\boldsymbol{\vec{\chi}^2}(100)$. At this point, the analytical function associated with the minimum $\chi^2$ is selected as the final output of the algorithm, referred to as $f_{\rm GA}(x)$:
\begin{align}\label{eq:fGAdef}
    \chi^2_{\rm min} = \min {\boldsymbol{\vec{\chi}^2}}(100), \longrightarrow f_{\rm GA}(x) = \mathcal{F}_{\chi^2_{\rm min}}(100).
\end{align}
Equation~\eqref{eq:fGAdef} can be read as the function coming out of the last generation that presents the minimum $\chi^2$, that is the function that best reproduce the input data points.

\subsection{Baseline GA reconstruction}\label{ssec:singleiter}
To demonstrate the algorithm in action and illustrate the proposed improvement, we follow the procedure described in \cite{Nesseris:2012}. First, we generate a set of 20 sparse mock data points $(x_i,\, y_i,\, \sigma_i)$ based on the model:
\begin{equation}\label{eq:testfunction}
    f(x;\,a,b,c) = a+(x-b)\cdot e^{-cx^2},
\end{equation}
where the parameters are $(a,b,c) = (0.75,0.25,0.10)$. The objective of GA is to generate a symbolic analytical function that reproduces the underlying $f(x)$ model only from data points.
 
\begin{figure}[h]
\centering
\includegraphics[width=\linewidth]{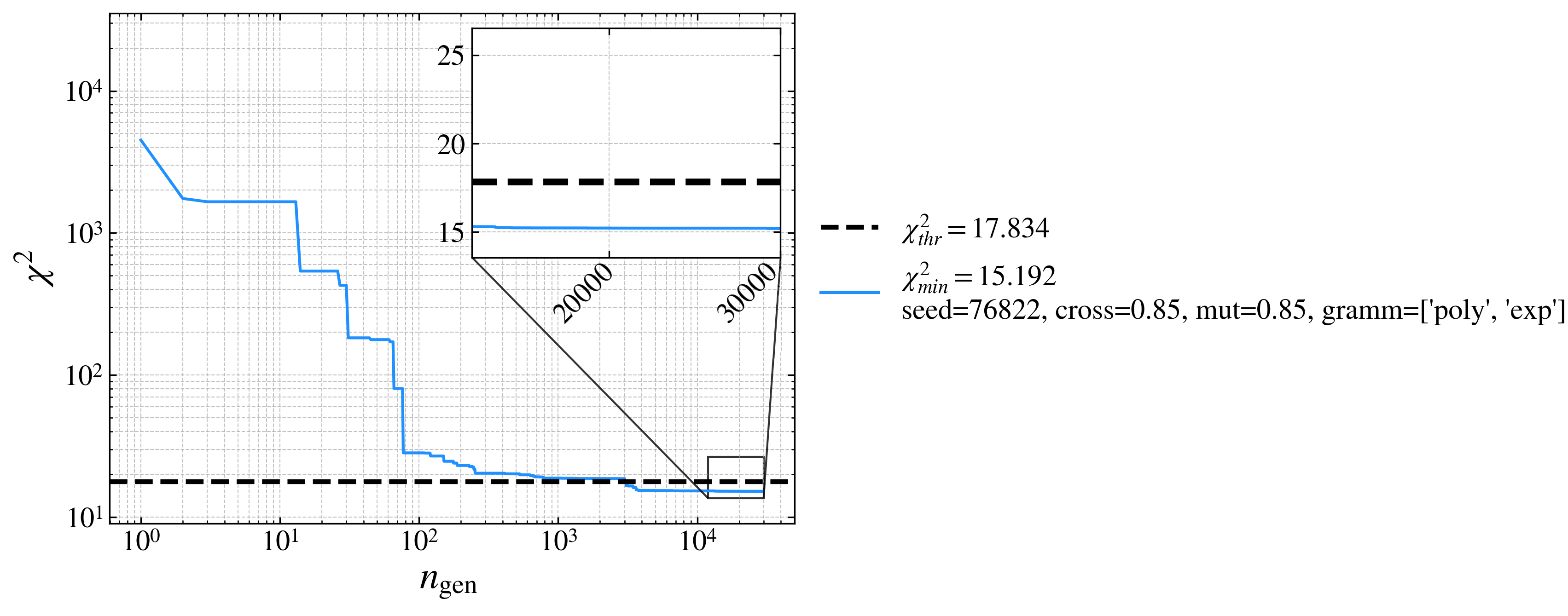}
\caption{Evolution of the GA $\chi^2$ with the number of generations $n_{\rm gen}$. The black dotted line is the $\chi^2_{\rm threshold}$, representing the $\chi^2$ we aim to overcome, obtained through standard model-dependent minimisation of the parametrised function. The blue curve represents the $\chi^2$ of the function generated by the GA. In the legend the hyperparameters setup of the GA are listed: the random seed is 76822, the probability of crossover and mutation is $85\%$, the grammar set for the generation is made of polynomial and exponential functions, the number of generations is $N_{\rm gen}=30000$, sufficient for $\chi^2_{\rm GA}$ to drop below $\chi^2_{\rm threshold}$.}
\label{fig:single_iteration_chi2}
\end{figure}

We selected a function that exhibits smooth behaviour and possesses a maximum on purpose, to demonstrate that this methodology is capable of detecting the features underlying the function we are attempting to reconstruct. Then, to establish a benchmark for comparison, we minimised $\chi^2_i(a,b,c)$ from Equation~\eqref{eq:chi2_uncorr} with respect to the parameters $(a,b,c)$. We refer to the obtained result as $\chi^2_{\rm threshold}$, representing the value obtained from a standard parametric analysis that we aim to surpass with a non-parametric algorithm. Next, we fix a set of hyperparameters for the GA and let the algorithm generate the best-fit function. Figure \ref{fig:single_iteration_chi2} shows the evolution of the $\chi^2$ of the GA-generated function against the number of generations $n_{\rm gen}$. Once the algorithm converges, if the curve goes below the threshold set by the standard parametric approach, it means we obtain an analytical function that fits data better than the one obtained from the model-dependent parameter minimisation.

\subsubsection{$\chi^2$ marginalisation}
Given that GA rely entirely on the estimator used to establish whether a function adequately represents data (in our case, the $\chi^2$), it is often useful to incorporate simple statistical techniques to refine the reconstruction process. 
A practical approach is to assume that the underlying model may differ from the GA reconstruction function by a constant $y_0$. Depending on the problem, this constant can act as an additive offset (e.g. an uncertain zero-point) or as a multiplicative normalisation (e.g. an unknown overall scale). 
\begin{itemize}
    \item \textbf{Additive offset:} $y(x) = y_0 + f_{\rm GA}(x)$;
    \item \textbf{Multiplicative offset:} $y(x) = y_0 \cdot f_{\rm GA}(x)$.
\end{itemize}
Physically, the additive offset is helpful when dealing with a calibration or systematic error that may add a constant independent of the overall reconstructed function. For example, when reconstructing the distance modulus $\mu(z)$ from SNe Ia, the absolute magnitude $M_B$ acts as a global calibration parameter. The relationship between the observed standardised apparent magnitude $m_B(z)$ and the theoretical model is:
\begin{equation}\label{eq:m_B}
    m_B(z) = \mu(z)+M_B = 5\log_{10}\left({\dfrac{D_L(z)}{\rm{Mpc}}}\right)+25+M_B.
\end{equation}
When working directly with pre-calibrated distance moduli $\mu(z)$, any uncertainty or variation in the absolute calibration $M_B$ translates into an additive offset (\cite{Riess:2021}) that we want to marginalise over. On the other side, the multiplicative offset is suited when dealing with a normalisation factor of the quantity we are interested in reconstructing. For the Hubble rate, and analogously the test function, we are more interested in the shape of $H(z)$, while the overall normalisation is set by the Hubble constant $H_0$, so it is natural to treat $H_0$ as a multiplicative offset.

We want to re-define the $\chi^2$ to take into account potential offsets (\cite{Arjona:2020, Nesseris:2007, Nesseris:2005, Nesseris:2004}). Inserting these expressions into the standard uncorrelated $\chi^2$ in Equation~\eqref{eq:chi2_uncorr}, the $\chi^2$ becomes a quadratic function of the single nuisance parameter $y_0$,
\begin{equation}
    \chi^2(y_0) = C\,y_0^2 - 2 B\,y_0 + A,
\end{equation}
where $A$, $B$ and $C$ depend only on the data and on the GA reconstruction $f_{\rm GA}$. Their explicit expressions are
\begin{itemize}
    \item \textbf{Additive offset:}
    \begin{equation}
        A \equiv \sum_{i=1}^{N_{\rm data}}
            \left(\frac{f_{\rm GA}(x_i)-y_i}{\sigma_i}\right)^2,\quad
        B \equiv \sum_{i=1}^{N_{\rm data}}
            \frac{y_i-f_{\rm GA}(x_i)}{\sigma_i^2},\quad
        C \equiv \sum_{i=1}^{N_{\rm data}}
            \frac{1}{\sigma_i^2};
    \end{equation}
    \item \textbf{Multiplicative offset:}
    \begin{equation}
        A \equiv \sum_{i=1}^{N_{\rm data}}
            \left(\frac{y_i}{\sigma_i}\right)^2,\quad
        B \equiv \sum_{i=1}^{N_{\rm data}}
            \frac{f_{\rm GA}(x_i)\,y_i}{\sigma_i^2},\quad
        C \equiv \sum_{i=1}^{N_{\rm data}}
            \left(\frac{f_{\rm GA}(x_i)}{\sigma_i}\right)^2.
    \end{equation}
\end{itemize}
Now, in order to reduce the impact of the offset, we want to marginalise $y_0$. Since $\chi^2(y_0)$ is quadratic, it can be minimised analytically with respect to $y_0$. One finds
\begin{equation}
    \dfrac{d\chi^2}{dy_0}\overset{!}{=}0, \quad\longrightarrow\quad y_0^{\rm best} = \dfrac{B}{C},
\end{equation}
and the corresponding minimum value of the $\chi^2$ is
\begin{equation}
    \chi^2_{\rm min} = A - \frac{B^2}{C}.
\end{equation}
This analytical minimisation allows us to use $\chi^2_{\rm min}$, a quantity now independent of the offset $y_0$, directly as the effective fitness function. This implies that the optimal constant offset is accounted for and does not need to be encoded in the chromosome. This way, GA do not have to waste generations simply to fix the offset, and for each $f_{\rm GA}(x)$ the best $y_0$ is found exactly and instantaneously. This speeds up convergence and makes the comparison between different grammars more robust, since functions with the same shape but different normalisations are not artificially penalised. Moreover, different grammars can easily generate different scales, marginalising with respect to an offset reduces the possibility of the GA randomly favouring reconstructions with a normalisation closer to data, rather than reconstructions with better behaviour and representation of the underlying model. From this point on, any $\chi^2$ appearing in the GA context will denote the marginalised quantity $\chi^2_{\rm min}$.

\subsubsection{Error analysis: path-integral}\label{ssec:pathintegral}
A limitation of GA is that they do not provide a direct method for estimating the uncertainties of the resulting analytical function. Furthermore, the fact that the algorithm samples the function space by randomly generating function combinations can complicate the interpretation of errors associated with the best-fit. An efficient and innovative approach is therefore required to extrapolate a function $\delta f(x)$ indicative of the uncertainty along the reconstructed function $f_{\rm GA}(x)$. While bootstrap Monte Carlo offers one solution \cite{Nesseris:2010}, we adopt the approximation of the error using the \textit{path-integral approach} from \cite{Nesseris:2012}. This approach is compatible with uncertainty estimation in Monte Carlo simulations \cite{Nesseris:2012}, where errors are determined by sampling the parameter space around the best-fit \cite{NumericalRecipes}. Finally, it provides a way of obtaining an analytical, smooth, uncertainty function which can be computed in every $x$, analogously to $f_{\rm GA}(x)$.

The fundamental idea is to treat the reconstructed function as the object of inference and to define a (formal) likelihood functional over the space of functions:
\begin{equation}\label{eq:likelihood}
\mathcal{L}[f] \propto \exp\left[-\frac{1}{2}\chi^2[f]\right],
\end{equation}
where $\chi^2[f]$ is the same goodness-of-fit functional estimator adopted in the GA. Equation \eqref{eq:likelihood} corresponds to the standard Gaussian likelihood, but unlike standard parametric cosmological analyses where the likelihood explores a finite dimensional parameter space, here the likelihood is formally defined as a functional over the dimensional space consisting of analytical functions generated by the GA grammar. Under the assumption of Gaussian measurement errors and, in the simplest case, uncorrelated data points, the likelihood is the product of the single-point likelihoods at the sampled $x_i$. In this framework, the uncertainty band is constructed by determining the probability that the true function lies within an uncertainty band of half-width $\delta f(x)$ around a chosen reference reconstruction $f_{\rm GA}(x)$:
\begin{equation}
f_{\rm GA}(x) - \delta f(x) \le f(x) \le f_{\rm GA}(x) + \delta f(x).
\end{equation}
The path-integral construction offers a practical way to approximate this probability without explicitly resampling the dataset; it is computationally less demanding than bootstrap Monte Carlo while generating comparable confidence regions. Concretely, for each data point $(x_i,\, y_i,\, \sigma_i)$ with statistically independent errors, this probability is obtained by integrating the Gaussian likelihood over the relevant interval, leading to a closed-form expression in terms of error functions:
\begin{equation}
CI(x_i,\delta f_i) = \frac{1}{2} \left[ \mathrm{erf}\left(\frac{\delta f_i + f_{\rm GA}(x_i) - y_i}{\sqrt{2}\sigma_i}\right) + \mathrm{erf}\left(\frac{\delta f_i - f_{\rm GA}(x_i) + y_i}{\sqrt{2}\sigma_i}\right) \right].
\end{equation}
The $1\sigma$ half-width $\delta f_i$ at each point is then obtained by numerically solving for the error that imposes the standard $68\%$ confidence level of a Gaussian distribution:
\begin{equation}CI(x_i,\delta f_i)\overset{!}{=}\mathrm{erf}\left(\frac{1}{\sqrt{2}}\right)\simeq 0.6827.\end{equation}

If one were to determine a separate $\delta f_i$ at each $x_i$, the resulting band would display strong point-to-point fluctuations driven by the noise in the data. To avoid this, we model $\delta f(x)$ as a smooth function (for example, a low-order polynomial $\delta f(x;\alpha, \beta, \gamma)=\alpha x^2+\beta x + \gamma$) and fix its parameters by minimizing a functional constructed from the confidence intervals over the full dataset:
\begin{equation}
\chi^2_{\rm CI}[\delta f] = \sum_{i=1}^{N_{\rm data}} \left[CI(x_i,\delta f(x_i)) - \mathrm{erf}\left(\frac{1}{\sqrt{2}}\right)\right]^2.
\end{equation}
In practice, we constrain the solution by requiring that the perturbed reconstructions $f_{\rm GA}(x)\pm\delta f(x)$ remain statistically consistent with the data, according to the same $\chi^2$ metric used in the original GA fit. In this way, the final band reflects the measurement-driven statistical uncertainty encoded in the likelihood. The result is a smooth estimate of the $1\sigma$ confidence region around the GA reconstruction, obtained without the need to repeatedly rerun the GA on many mock datasets.

The path-integral result can be extended to the case where data points are correlated \cite{Arjona:2020}. Following \cite{Nesseris:2012}, one can account for correlations and incorporate the covariance matrix, thus modifying the $\chi^2$ entering the GA to
\begin{equation}\label{eq:correlatedchi2}
    \chi^2 = \sum_{i,j}^{N_{\rm data}} \left(y_i-f(x_i)\right)\, C^{-1}_{ij}\left(y_j-f(x_j)\right),
\end{equation}
where $C_{ij}$ is the covariance matrix. For the final uncertainty estimate, the derivation follows the uncorrelated case but requires rotating the data into a basis in which they are uncorrelated (see Section II.C.2 of \cite{Nesseris:2012}). In this correlated framework, one obtains an analogous polynomial $\delta f_{\rm PI}(x)$ that now incorporates the correlations between points.

In what follows, we will denote this function simply as $\delta f_{\rm PI}(x)$, emphasizing that it represents the measurement-driven uncertainty derived via the path-integral method.

\subsection{Ensemble of hyperparameter models}
A sensitive point of GA is their intrinsic stochasticity and the choice of hyperparameters. The final function $f_{GA}(x)$ may differ considerably depending on the selected grammar set or on values of crossover and mutation probabilities. However, since the hyperparameter space contains elements that cannot be easily marginalised (such as the \texttt{grammar}), a standard marginalisation is not feasible. In addition, even if all hyperparameters are fixed, different random seeds can still produce substantially different best-fit functions. 

\begin{figure}[h]
\centering
\includegraphics[width=\linewidth]{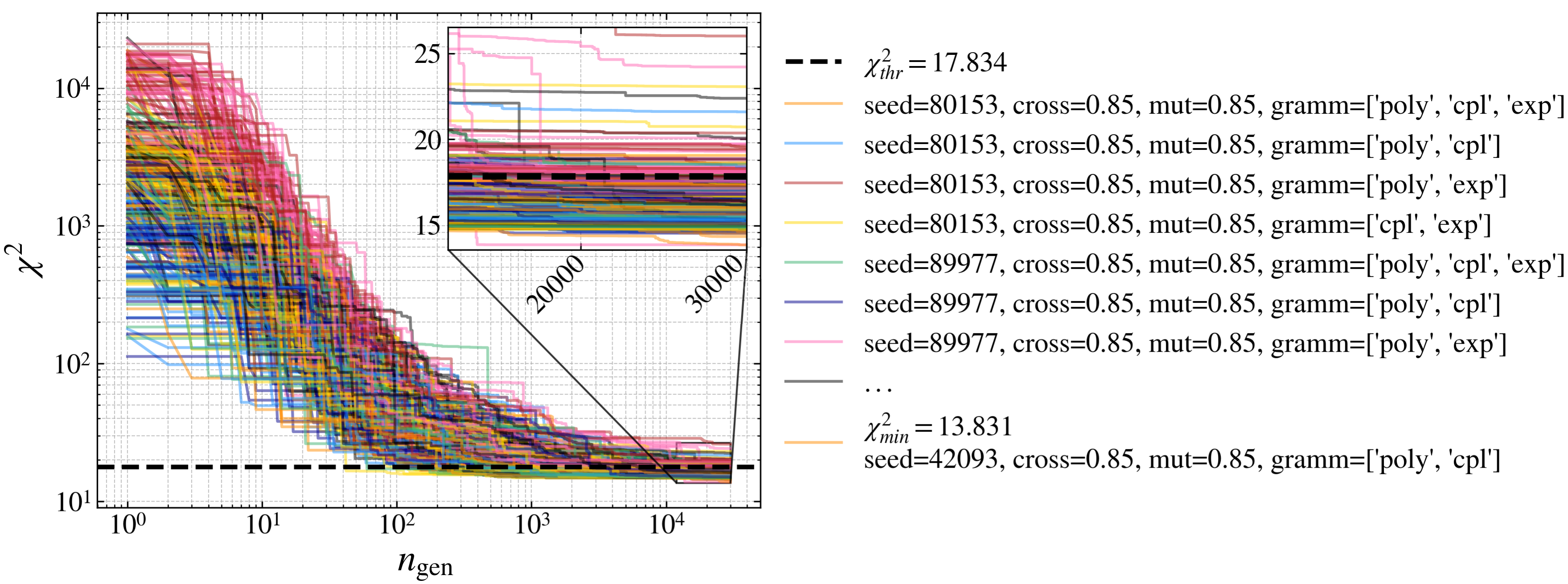}
\caption{Evolution of the GA $\chi^2$ with the number of generations $n_{\rm gen}$. The black dotted line is the $\chi^2_{\rm threshold}$, representing the $\chi^2$ we aim to overcome, obtained through standard model-dependent minimisation of the parametrised function. The coloured curves represent the $\chi^2$ of the functions generated by the GA for different configurations of the hyperparameters. In the legend each hyperparameters setup of the GA is listed.}
\label{fig:many_iterations_chi2}
\end{figure}

In order to identify the best hyperparameter setup, the typical strategy is to repeatedly run the GA with $N_{\rm conf}$ different hyperparameter choices and track the corresponding evolution illustrated in Figure \ref{fig:many_iterations_chi2}.

In line with Section ~\ref{ssec:singleiter}, once the final generation is reached, the only new ingredient is that we now have a set of $N_{\rm conf}$ best-fit functions. The straightforward GA prescription would be to select the single overall best-fit function, i.e. the one with the absolute minimum $\chi^2$. However, this result would still represent just one particularly lucky hyperparameter configuration. The selected function does not interact with the other candidates and does not reflect the impact of hyperparameter variations. Furthermore, we cannot guarantee that choosing a different endpoint for $N_{\rm gen}$ would still favour the same optimal function (as suggested by Figure \ref{fig:many_iterations_chi2}, where stopping earlier could return the best-fit of a different configuration). Indeed, Figure~\ref{fig:many_iterations_chi2} provides a clear convergence diagnostic for the optimisation process: it shows how separate GA runs may become stuck at different $\chi^2$ plateaus, effectively trapping the algorithm in local minima. Since focusing only on the global minimum in $\chi^2$ limits the potential of GA, we instead adopt a \textit{model-averaging approach}. This strategy reduces the effects of stochasticity while incorporating multiple models across the hyperparameter space, offering greater stability for the resulting function, especially with respect to its derivatives.

\subsubsection{Marginalisation over the ensemble (\texttt{GAME})}\label{ssec:game}
\begin{figure}[h]
\centering
\includegraphics[width=\linewidth]{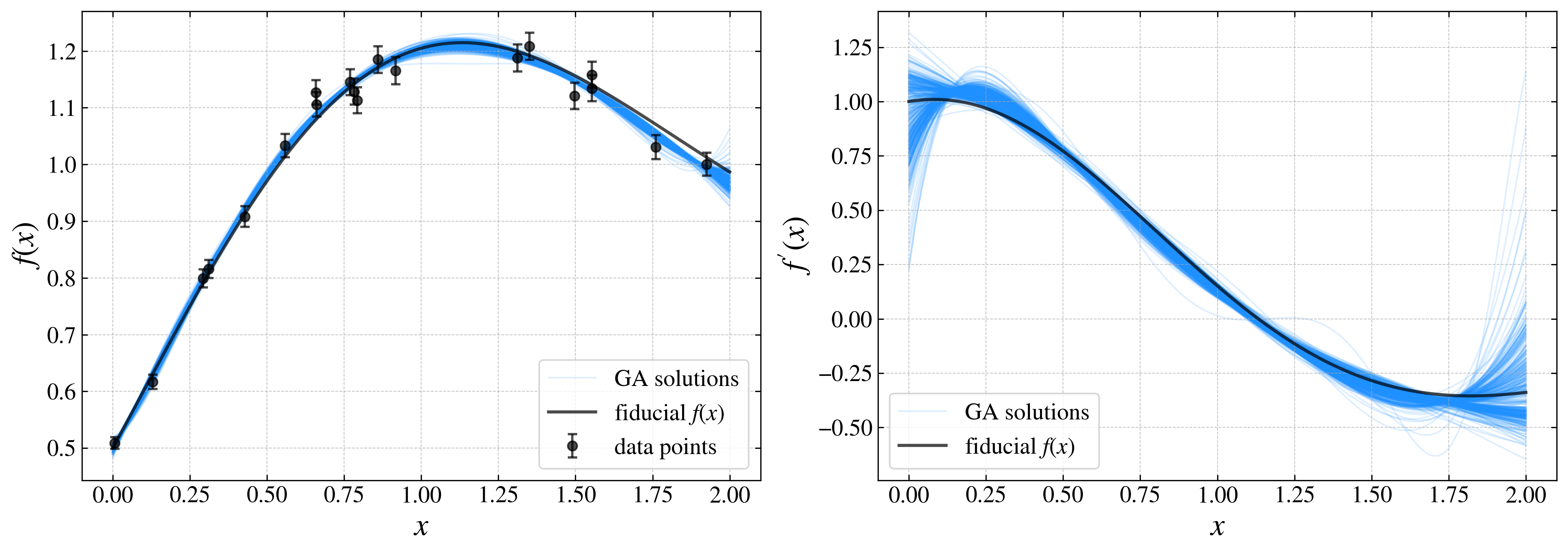}
\caption{Ensemble of reconstructions obtained by running the GA across multiple hyperparameter configurations. \textit{Left:} Reconstructed functions $f_{j,\rm GA}(x)$ (blue curves) compared with the fiducial model we are trying to reproduce (black line) and the mock data points with uncertainties generated on the fiducial model. \textit{Right:} Corresponding derivatives $f'_{j,\rm GA}(x)$ for the same ensemble, highlighting how the spread between GA solutions grows when taking derivatives, especially near the boundaries. The collection of curves motivates a model-averaging treatment to obtain a more stable reconstructed function and derivative.}
\label{fig:spaghetti_poly}
\end{figure}

We face $N_{\rm conf}$ different configurations, and for each configuration we obtain one best-fit function $f_{j,\rm GA}$, forming a set of $N_{\rm conf}$ candidate functions (Figure \ref{fig:spaghetti_poly}) that we wish to combine. Since the GA output is analytical, we compute a weighted average of analytic reconstructions, which serves as a representative combination of multiple GA solutions. This procedure incorporates all possible contributions across the hyperparameter space, including the stochasticity induced by the choice of random seed.

To perform averaging, we must define an estimator. A first attempt might consider the minimum $\chi^2$ and define corresponding $N_{\rm conf}$ normalised weights $w_j$ with an inverse-square or an exponential cut-off:
\begin{equation}\label{eq:likeliweights}
    w_j = \dfrac{\exp{\left[- \dfrac{1}{2}\left(\chi^2_j-\chi_{\rm min}^2\right)\right]}}{\displaystyle\sum_{k=1}^ {N_{\rm conf}} \exp{\left[- \dfrac{1}{2}\left(\chi^2_k-\chi_{\rm min}^2\right)\right]}}, \qquad \text{with}\quad \sum_{j=1}^{N_{\rm conf}} w_j = 1.
\end{equation}
However, this does not address the smoothness of the final function nor the fact that we are attempting to reproduce a physical quantity from statistical assumptions. A very low $\chi^2$ does not guarantee that the function faithfully reproduces the underlying physical model; often, it indicates overfitting, which becomes evident only when considering derivatives. Therefore, using $\chi^2$ alone is insufficient. We define a new estimator to weigh the average and compute it for each of the $N_{\rm conf}$ functions:
\begin{equation}\label{eq:estimator}
    S_j(\lambda) = \chi^2_j + \lambda R_j,\qquad \text{with}\quad R_j = \int \left|f^{''}_j(x)\right|^2\, dx,
\end{equation}
where $R_j$ is the \emph{roughness} term  \cite{Beyhum:2024}, quantifying the smoothness of the first derivative. In our implementation, this is computed numerically on the same grid used for plotting and likelihood evaluations. $\lambda$ is a regularisation parameter, an arbitrary real number quantifying the relative importance of the roughness term. The lower its value, the more weight is given to the $\chi^2$; conversely, the greater it is, the more the smoothness is prioritised over the data fit. Ideally, $\lambda$ is fixed via a trade-off between smoothness and representativeness. We employ the \emph{L-curve method} \cite{Hansen:2001, Calvetti:2004}, optimising the choice of regularisation parameter based on the log-log plot of the two quantities $\chi^2$ and $R$. Note that we also tried to marginalise over $\lambda$ by treating it as an additional hyperparameter. Although the result was consistent with the L-curve technique, this procedure was significantly slower; therefore, we chose to retain the L-curve method for reasons of computational efficiency.

\begin{figure}[h]
\centering
\includegraphics[width=12cm]{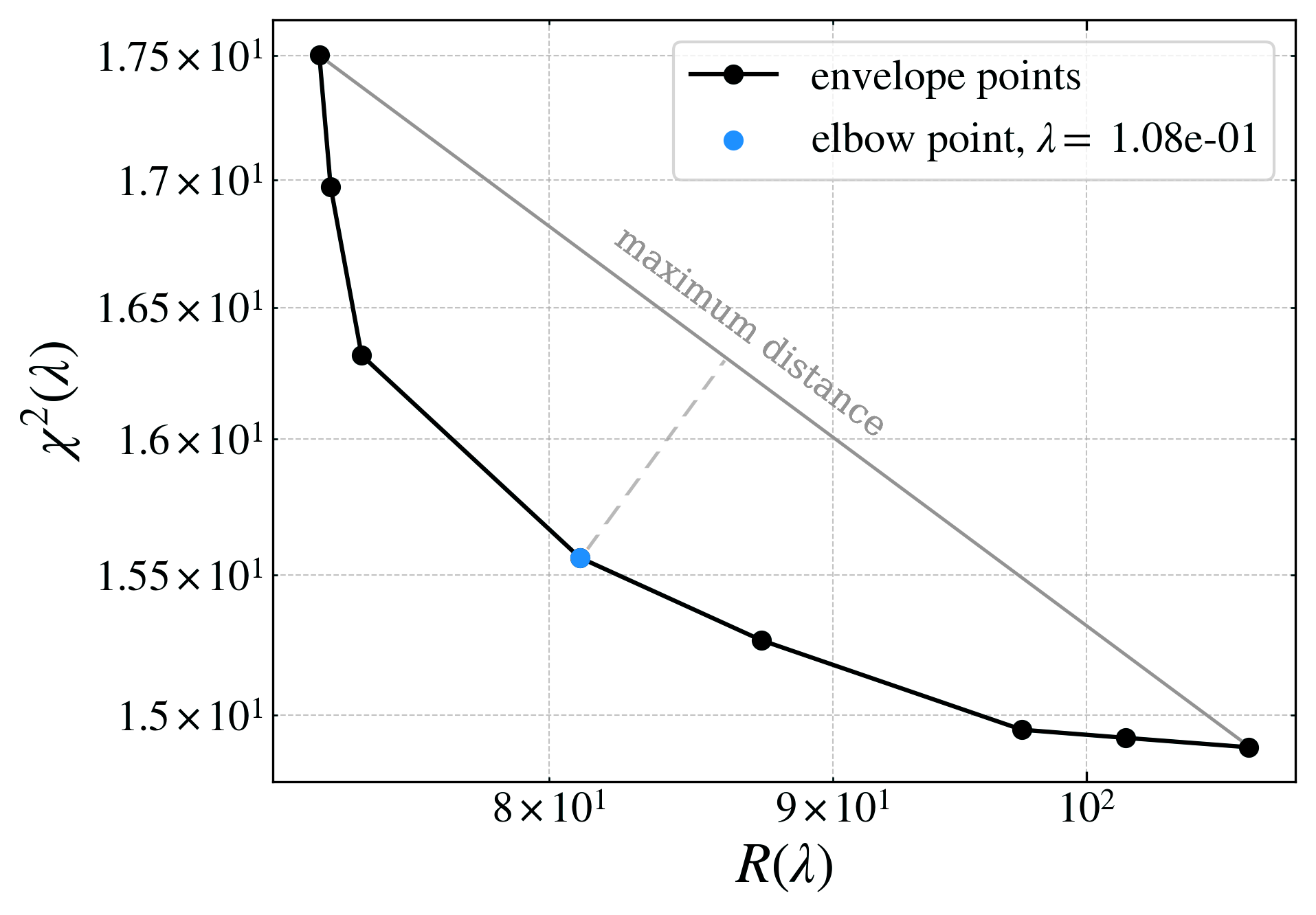}
\caption{L-curve used to identify the optimal regularization parameter $\lambda$ for the estimator $S_j(\lambda) = \chi_j^2 + \lambda R_j$. Black markers represent the lower-envelope points $(R_{j_\star}, \chi^2_{j_\star})$ obtained by scanning the $\lambda$ grid and identifying the specific configuration $j_\star$ that minimizes $S_j$ at each step. These points represent the active subset of the $N_{\rm conf}$ ensemble; their limited number indicates that many $\lambda$ values map to the same optimal model $j_\star$. The blue marker identifies the elbow point, defined by the maximum curvature in the $\log R$-$\log \chi^2$ plane, where the trade-off between goodness-of-fit ($\chi^2$) and smoothness ($R$) is optimized. This determines $\lambda_{\rm elbow}$, which is subsequently used to calculate the exponential weights $w_j$ for the final model-averaging procedure.}
\label{fig:elbow_point}
\end{figure}

To identify the optimal trade-off between $\chi^2$ and $R$, we construct a cost matrix by evaluating $S_j(\lambda)$ over a dense discrete grid of $\lambda\in\left[10^{-5},10^5\right]$ for all $N_{\rm conf}$ models:
\begin{equation}\label{matrix:S}
\begin{matrix}
  & \xrightarrow{\hspace*{2.5cm} \displaystyle \lambda \hspace*{2.5cm}} \\[5pt]
  j \left\downarrow \vphantom{\begin{pmatrix} 
      S_1 \\ \vdots \\ S_j \\ \vdots \\ S_N 
  \end{pmatrix}} \right. 
  & 
  \begin{pmatrix}
    S_1(\lambda_1) & \dots & S_1(\lambda_i) & \dots & S_1(\lambda_{\text{fin}}) \\
    \vdots & \ddots & \vdots & \ddots & \vdots \\
    S_j(\lambda_1) & \dots & S_j(\lambda_i) & \dots & S_j(\lambda_{\text{fin}}) \\
    \vdots & \ddots & \vdots & \ddots & \vdots \\
    S_{N_{\text{conf}}}(\lambda_1) & \dots & S_{N_{\text{conf}}}(\lambda_i) & \dots & S_{N_{\text{conf}}}(\lambda_{\text{fin}})
  \end{pmatrix}
\end{matrix}
\end{equation}
From this matrix we are able to extract the lower envelope of the model set. In practice, for each column (representing a specific $\lambda_i$), we identify the row element (representing a specific hyperparameter configuration $j_\star$) corresponding to the minimum values of $S_j(\lambda_i)$:
\begin{equation}
j_\star(\lambda_i)=\arg\min_{j\in{1,\dots,N_{\rm conf}}} S_j(\lambda_i).
\end{equation}
By scanning the $\lambda$ grid results of $S_j$ in the matrix \eqref{matrix:S}, and then locating $S_{j_\star}(\lambda_i)$, it is possible to identify and collect for each $\lambda_i$ the corresponding pair $\chi^2_{j_\star}$ and $R_{j_\star}$ associated with the minimum, i.e., the optimal pair configuration that produces the best $S$ at a given $\lambda$:
\begin{equation}
S_{\rm min}(\lambda)=S_{j_\star}(\lambda), \quad\longrightarrow\quad \left(R_{j_\star(\lambda)}, \chi^2_{j_\star(\lambda)}\right).
\end{equation}
Note that many different values of $\lambda_i$ can correspond to the same optimal model $j_\star$ (i.e., as we vary $\lambda$, we may keep obtaining the same index $j_\star$ and the same corresponding minimum value $S_{j_\star}$). The reason is that, with $\chi^2_j$ fixed, minimising the sum with $\lambda R_j$ often produces only a limited number of distinct $S_{\rm min}$ values, making the outcome quite sensitive to the step size chosen for $\lambda$. As a result, the full scan effectively identifies only a relatively small subset of configurations that are effectively \emph{active} in the trade-off. This process filters the ensemble down to the small subset of models that are optimal for at least one value of $\lambda$. The pairs $\left(R_{j_\star},\chi_{j_\star}^2\right)$ corresponding to this subset are then plotted to form the L-curve in Figure \ref{fig:elbow_point}.

The L-curve method involves choosing the regularisation parameter $\lambda$ at the point where the curve exhibits the largest bending in the $\log$-$\log$ space (the so-called \emph{elbow}) \cite{Hansen:2001, Calvetti:2004}. This point represents the optimal trade-off, marking the transition between the regime driven mainly by $\chi^2$ (small $\lambda$), where the model fits the noise, and the regime dominated by the regularity constraint (large $\lambda$), where the actual signal begins to be smoothed. It serves as a criterion to select a regularisation parameter introduced without any prior. In practice, after collecting and ordering the envelope points by increasing $R$, we work in $\log$-$\log$ space and identify the elbow as the point of maximum distance from the straight line connecting the two endpoints of the envelope.

This produces a characteristic configuration $j_{\rm elbow}$ and an associated representative value $\lambda_{\rm elbow}$, chosen within the range of $\lambda$ for which $j_{\rm elbow}$ minimises $S_j(\lambda)$. Once $\lambda_{\rm elbow}$ is fixed, we define the weights $w_j$ through $S_j$:
\begin{equation}
w_j(\lambda_{\rm elbow})=
\dfrac{\exp{\left[- \dfrac{1}{2}\left(S_j(\lambda_{\rm elbow})-S_{\rm min}(\lambda_{\rm elbow})\right)\right]}}
{\displaystyle\sum_{k=1}^{N_{\rm conf}}\exp{\left[- \dfrac{1}{2}\left(S_k(\lambda_{\rm elbow})-S_{\rm min}(\lambda_{\rm elbow})\right)\right]}},
\end{equation}
which corresponds to an exponential cut-off driven by the combined estimator $S$. We can interpret this estimator not only as a likelihood ($\chi^2$) but as a posterior with regularisation prior ($\lambda R$).

The averaged analytical function is then computed as a weighted sum of the symbolic $N_{\rm conf}$ best-fit solutions of the GA:
\begin{equation}\label{eq:weightedGA}
f_{\rm GAME}(x)=\sum_{j=1}^{N_{\rm conf}} w_j(\lambda_{\rm elbow})\cdot f_{j,\rm GA}(x).
\end{equation}

\subsubsection{Error analysis: ensembling and total uncertainty}\label{ssec:totalerror}
The model-averaging described above mitigates the intrinsic stochasticity and the impact of hyperparameters setup of the GA. However, working with an averaged reconstruction requires an upgrade in the uncertainty estimation: besides the measurement-driven statistical uncertainty on the resulting function ($\delta f_{\rm PI}(x)$), we must account for an additional configuration-driven spread, that is the dispersion among the different GA best-fit functions entering the ensemble ($\sigma_{\rm ens}$).

For the former, we quantify the statistical uncertainty around the reconstructed function using the path-integral approach described in Section \ref{ssec:pathintegral}. In the averaged scenario, we adopt the weighted-mean reconstruction $f_{\rm GAME}(x)$ and generate a smooth error band by imposing the $1\sigma$ confidence interval condition. The function $\delta f_{\rm PI}(x)$ encodes the uncertainty implied by the likelihood functional $\mathcal{L}[f]\propto \exp[-\chi^2[f]/2]$.

For the latter, we consider the set of best-fit GA reconstructions $f_{j,\rm GA}$ and the associated weights $w_j$, evaluated at $\lambda_{\rm elbow}$, obtaining the weighted-averaged analytical function given in Equation \eqref{eq:weightedGA}. At each $x$, we define a weighted ensemble variance as a quantitative measure of the remaining configuration uncertainty \cite{Burnham:2002}:
\begin{equation}
\sigma_{\rm ens}^2(x)
= \sum_{j=1}^{N_{\rm conf}} w_j^2(\lambda_{\rm elbow})\Big(f_{j,\rm GA}(x)-f_{\rm GAME}(x)\Big)^2,
\label{eq:sigma_ens}
\end{equation}
which quantifies how much the reconstructions fluctuate around the averaged solution when scanning the hyperparameters space. Note that $\sigma_{\rm ens}(x)$ is tied to the stability of the reconstruction procedure itself rather than the measurement errors of a single run.

Finally, we combine the two contributions into a single effective $1\sigma$ band. Since the measurement noise (captured by $\delta f_{\rm PI}$) and the stochastic nature of the GA (quantified by $\sigma_{\rm ens})$ are uncorrelated processes, we treat them as independent sources of error and therefore sum their variances in quadrature:
\begin{equation}
\sigma_{\rm tot}^2(x) = \delta f_{\rm PI}^2(x) + \sigma_{\rm ens}^2(x), \quad\longrightarrow\quad f_{\rm GAME}(x)\pm \sigma_{\rm tot}(x),
\label{eq:sigma_tot}
\end{equation}
This provides a conservative and numerically stable error estimate. Here, $\delta f_{\rm PI}(x)$ is computed for the final averaged reconstruction $f_{\rm GAME}(x)$ (representing the statistical measurement uncertainty driven by the data likelihood), independent of the $N_{\rm conf}$ explored. In contrast, $\sigma_{\rm ens}(x)$ inflates the error band in regions where different GA configurations produce significantly different functional behaviours, which is particularly relevant for derivatives as seen in Figure \ref{fig:spaghetti_poly}. It is important to specify how the combined uncertainty band $\sigma_{tot}(x)$ should be interpreted statistically. This band represents a hybrid uncertainty estimate: the first component, $\delta f_{\rm PI}(x)$, provides a probabilistic estimate of the measurement-driven statistical error derived from a Gaussian likelihood functional, while the second contribution, $\sigma_{\rm ens}(x)$, is an empirical, heuristic measure of the systematic uncertainty associated with variations in the algorithm's hyperparameter configurations. By summing these independent contributions in quadrature, $\sigma_{\rm tot}$ acts as an effective uncertainty band that conservatively accounts for both data noise and the spread due to the algorithm.

The same logic applies to derived quantities such as $f'(x)$. We compute the analytical derivative for each model, then build the weighted-averaged derivative:
\begin{equation}
f'_{\rm GAME}(x)=\sum_{j=1}^{N_{\rm conf}} w_j(\lambda_{\rm elbow})\cdot f'_{j,\rm GA}(x),
\end{equation}
and define the corresponding weighted ensemble variance for the derivative:
\begin{equation}
\sigma_{\rm ens}^2\!\left[f'(x)\right]
= \sum_{j=1}^{N_{\rm conf}} w_j^2(\lambda_{\rm elbow})\Big(f'_{j,\rm GA}(x)-f'_{\rm GAME}(x)\Big)^2.
\end{equation}
For the path-integral contribution $\delta f_{\rm PI}$, we propagate the uncertainty by differentiating the reconstruction band. This assumes that the variation operator $\delta$ commutes with the derivative:
\begin{equation}
    \delta\left(\dfrac{df}{dx}\right)=\dfrac{d}{dx}\delta f(x).
\end{equation}
While this commutation is not generally true for arbitrary distributions, it is consistent with the Gaussian likelihood approximation ($\mathcal{L}\propto e^{-\chi^2/2}$) employed in the path-integral framework (Section \ref{ssec:pathintegral}). Under this approximation, the perturbations are treated as linear fluctuations around the best-fit, allowing for standard error propagation \cite{Arjona:2020}.

The final $1\sigma$ uncertainty on the derivative of the \texttt{GAME} function is obtained analogously to Equation \eqref{eq:sigma_tot}:
\begin{equation}
\sigma_{\rm tot}^2\!\left[f'(x)\right]
= \delta f_{\rm PI}'^{\,2}(x) + \sigma_{\rm ens}^2\!\left[f'(x)\right],
\quad\longrightarrow\quad
f'_{\rm GAME}(x)\pm \sigma_{\rm tot}\!\left[f'(x)\right].
\end{equation}

\subsubsection{Improvements in the reconstruction}\label{ssec:f_mock}
Finally, we fix the different $N_{\rm conf}$ hyperparameter configurations (Appendix \ref{app:ga_hyperparams}) and feed the algorithm with the mock data. It generates a set of candidates from which we identify the baseline best-fit GA function $f_{\rm GA}(x)$ (corresponding to the minimum $\chi^2$ between all the configurations) and compute the \texttt{GAME} averaged function $f_{\rm  GAME}(x)$ with its associated total uncertainty. We then compare the reconstructions obtained via the two methodologies. We emphasise that for all comparisons presented in this work, the baseline GA and the \texttt{GAME} reconstructions are obtained from the same ensemble $N_{\rm conf}$ runs, using identical hyperparameters for each run and the same data inputs. The only distinction between the two methods lies in the post-processing applied after the ensemble has been generated: the baseline GA reports the single global minimum $\chi^2$ solution together with its path-integral uncertainty, whereas \texttt{GAME} outputs the weighted-average function given in Equation \eqref{eq:weightedGA} along with the total uncertainty defined in Equation \eqref{eq:sigma_tot}.
\begin{figure}[h]
\centering
\includegraphics[width=\linewidth]{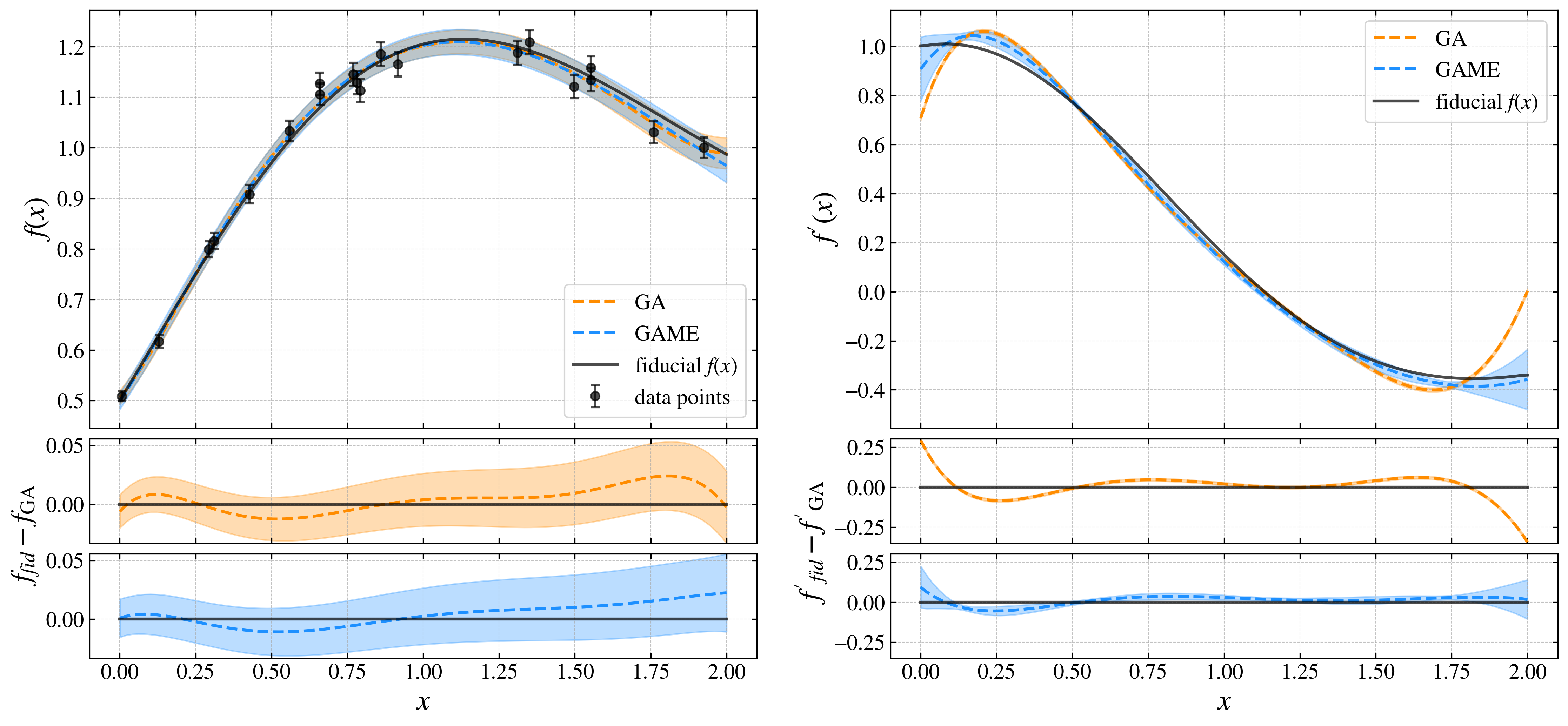}
\caption{Comparison of function reconstruction and derivative estimation between the baseline GA and \texttt{GAME} methodologies. The bottom sub-panels display the residuals for the GA and \texttt{GAME} methods, respectively. \textit{Left:} Reconstruction of the function $f(x)$. The solid black line indicates the underlying fiducial function (Equation~\eqref{eq:testfunction}) we are trying to reproduce from the mock dataset. The orange dashed line shows the baseline best-fit GA result, while the blue dashed line represents the \texttt{GAME} averaged reconstruction. Shaded regions indicate $1\sigma$ confidence intervals: a purely statistical confidence interval ($\delta f_{\rm PI}$) for the baseline GA, and a hybrid band combining the statistical uncertainty with the configuration ensemble variance ($\sigma_{\rm ens}$) for \texttt{GAME} (Section~\ref{ssec:totalerror}). \textit{Right:} Derivation of the derivative $f'(x)$. While the reconstructed functions $f_{\rm GA}$ and $f_{\rm GAME}$ are comparable, the derivative estimation highlights the stability of the \texttt{GAME} averaging at the boundaries ($x \approx 0$ and $x \approx 2$), where the baseline GA deviates significantly from the fiducial line.}
\label{fig:GAME_poly}
\end{figure}

In Figure \ref{fig:GAME_poly}, we observe that for the reconstruction of $f(x)$, both methodologies generate analytical functions that are compatible at the $1\sigma$ level. However, \texttt{GAME} reconstruction tends to be more stable and resistant to overfitting. To rigorously quantify this improvement, we introduce the cumulative statistic $\tilde\chi^2_{\rm tot} = \sum_{x} \left[\left(f_{\rm fid}(x)-f_{\rm GA}(x)\right)/\sigma_{\rm tot}(x)\right]^2$. This estimator serves as a metric to evaluate and compare how well the reconstructions reproduce the underlying target model\footnote{The tilde $\sim$ is introduced to distinguish this model-function $\chi^2$ from the data-function $\chi^2$ used within the GA (Equation~\eqref{eq:chi2_uncorr}).} $f(x)$.
\begin{figure}[h]
\centering
\includegraphics[width=\linewidth]{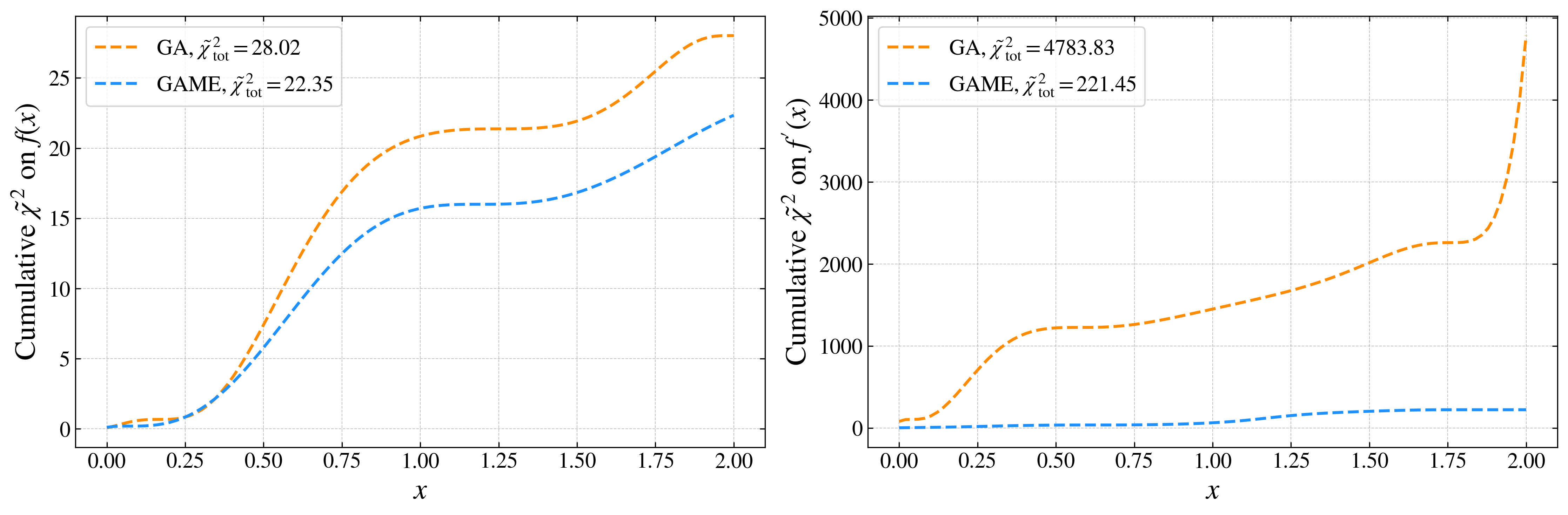}
\caption{Cumulative performance of the reconstruction methodologies against the true fiducial model. The panel show the cumulative $\tilde\chi^2_{\rm tot}$ as a function of $x$, calculated between the reconstructed function and the underlying target model, normalised by the total uncertainty $\sigma_{\rm tot}$. \textit{Left:} Cumulative $\tilde\chi^2_{\rm tot}$ for the function $f(x)$. The blue line (\texttt{GAME} approach) reaches a final value of $\tilde\chi^2_{\rm tot}\approx22$, representing a $\sim 20\%$ improvement over the orange line (baseline GA best-fit) $\tilde\chi^2_{\rm tot}\approx 28$. \textit{Right:} Cumulative $\tilde\chi^2_{\rm tot}$for the derivative $f'(x)$. Here the difference is drastic: the baseline GA accumulates a massive error at the boundaries due to instability ($\tilde\chi^2_{\rm tot}\approx 4783$), while \texttt{GAME} maintains stability throughout ($\tilde\chi^2_{\rm tot}\approx221$) resulting in a $\sim 95\%$ improvement in accuracy.}
\label{fig:cumulative_chi2}
\end{figure}
In Figure~\ref{fig:cumulative_chi2}, we observe the evolution of the cumulative $\tilde\chi^2_{\rm tot}$ for both the function and it derivative. While the improvement for $f(x)$ is visually subtle in the reconstruction plots, the cumulative estimator reveals a quantitative improvement of $\sim 20\%$.

For the derivative $f'(x)$, the advantage of the \texttt{GAME} approach is evident. As seen in the right panel of Figure~\ref{fig:GAME_poly}, the differentiated GA best-fit is inaccurate at both low and high values of $x$ despite having the global minimum $\chi^2$ on the data. Instead, by applying \texttt{GAME} and weighting the ensemble with an estimator sensitive to roughness, makes the derivative significantly more stable, with an uncertainty that accounts for major deviations (in this case, at low and high $x$ values), and achieving an improvement of $\sim 95\%$.

This test confirms that the \texttt{GAME} methodology has the potential to reconstruct generic functions and derived quantities in a non-parametric and model-independent manner. We proceed by applying \texttt{GAME} to cosmological quantities and real data to assess the quality of such reconstructions and their ability to constrain cosmological parameters.

\section{Application to cosmological quantities}\label{sec:CosmoApp}
\subsection{Dark energy equation of state}
We aim to test the standard $\Lambda$CDM model through dark energy, which is not a directly observable quantity. One can study its equation of state $w$ using direct observations of background quantities such as the luminosity distance of SNe Ia $D_L(z)$, or the expansion rate of the universe, $H(z)$. From $H(z)$, $w$ can be derived following parametric or non-parametric approaches \citep{Huterer:2017}.
We assume the universe is well described by an FLRW metric and consider a Friedmann equation for a late-time flat universe containing only matter and dark energy (neglecting radiation at low redshifts):
\begin{equation}
    \dfrac{H(z)^2}{H_0^2}=\Omega_{m,0}(1+z)^3+\Omega_{\rm de}(z),\qquad \text{with} \quad\Omega_{\rm de}(z) = \rho_{\rm de}(z)\dfrac{8\pi G}{3H(z)^2}.
\end{equation}
Assuming dark energy acts as a perfect fluid ($p_{\rm de}\equiv w(z)\rho_{\rm de}$), and as a non-interacting fluid ($\dot{\rho}_{\rm de}+3H(1+w)\rho_{\rm de}=0$), we obtain \cite{Bogdanos:2009}:
\begin{equation}\label{eq:de_eos}
    w(z) = -1+\dfrac{1}{3}(1+z)\dfrac{d\ln{\left(\dfrac{H(z)^2}{H_0^2}-\Omega_{m,0}\left(1+z\right)^3\right)}}{dz}.
\end{equation}
This expression links the evolving dark energy equation of state to background evolution quantities. By reconstructing $H(z)$ from data and assuming a prior value for $\Omega_{m,0}$, we can perform a consistency check with the expected $\Lambda$CDM value ($w=-1$).

However, Equation~\eqref{eq:de_eos} is derived under the assumptions of a flat FLRW geometry and the GR theory of gravity. If any of these assumptions is violated ($\Omega_k\ne 0$ or modified gravity affecting Friedmann equations), then the reconstructed $w(z)$ ceases to measure the physical properties of the dark energy fluid alone. Instead, it absorbs the effects of curvature and modified gravity, making them mathematically indistinguishable from a change in the dark energy equation of state.

Within this framework, any deviation from $w=-1$ can imply one of these physical scenarios \cite{Perenon:2022}:
\begin{enumerate}
    \item \textbf{Dynamical dark energy:} the fluid is not a vacuum energy density but a dynamical field with a time-evolving equation of state \cite{Copeland:2006}.
    \item \textbf{Interacting dark energy:} our derivation assumes the continuity equation. If there is an energy exchange between dark matter and dark energy, this interaction can mimic an effective equation of state $w(z)\ne-1$ even if the fluid itself is a cosmological constant \cite{Wang:2016}.
    \item \textbf{Matter density parameter bias:} the reconstruction of $w(z)$ is mathematically degenerate with the matter density parameter. An incorrect prior on $\Omega_{m,0}$ will induce a spurious evolution in $w(z)$ to compensate for the density mismatch \cite{Sahni:2008}.
\end{enumerate}

In order to disentangle the dark energy sector from the systematic parameter bias (scenario 3), we frame the current observational landscape by computing $w(z)$ using two distinct priors on $\Omega_{m,0}$:

\begin{itemize}\label{list:Omegampriors}
\item \textbf{Planck 2018} \cite{Planck:2018}: $\Omega_{m,0} = 0.315 \pm 0.007$, representing the CMB constraints.
\item \textbf{DESI DR1} \cite{DESI:2024}: $\Omega_{m,0} = 0.295 \pm 0.015$, representing recent constraints from BAO.
\end{itemize}

\subsection{Testing $\Lambda$CDM with Cosmic Chronometers}
To derive $w(z)$, we must reconstruct $H(z)$. A known issue is that any possible data relating the two quantities must be differentiated at least once. Distance data, such as SNe Ia, require two derivatives \cite{Copeland:2006}, greatly enhancing the error on $w(z)$. Following \cite{Perenon:2022}, we utilise direct measurements of $H(z)$ from CC. Although these observations are sparser and much more affected by systematics, they require only one derivative (see Equation \eqref{eq:de_eos}), making them good candidates to apply \texttt{GAME} methodology. CC are based on massive, passively evolving early-type galaxies that act as standard clocks. By measuring the differential age evolution $dt$ between two such galaxies separated by a small redshift interval $dz$, one can directly determine the expansion rate via the relation $H(z) = - \frac{1}{1+z} \frac{dz}{dt}$. Particularly, the dataset we have used comes from Table 1 of \cite{Favale:2023lnp}, at which we have added a recent measurement from \cite{Tomasetti:2023}. Unlike luminosity distance, this method measures the Hubble rate directly, allowing for a determination of the expansion history that does not rely on integrations over the equation of state $w(z)$ \cite{Jimenez:2002}. 

\subsubsection{Reconstruction of $H(z)$}\label{ssec:H_real}

We first fix the hyperparameters  (Appendix \ref{app:ga_hyperparams}), then we feed the GA with CC data to reconstruct $H(z)$. We consider a dataset of measurements at $z\lesssim 2$. Even though some of these data points are known to be correlated \cite{Moresco:2020}, we treat the sample as uncorrelated to perform an additional robustness test, with the objective of remaining as agnostic as possible in order to test the goodness of the methodology. Including correlations injects extra information and, as seen in Section \ref{ssec:pathintegral}, correlated data just impact the value of the $\chi^2$ entering the GA (Equation~\eqref{eq:correlatedchi2}). In case we wanted to build a covariance matrix for the CC $H(z)$ measurements, this could be done following the procedure described in \cite{Moresco:2020}.

\begin{figure}[h]
\centering
\includegraphics[width=\linewidth]{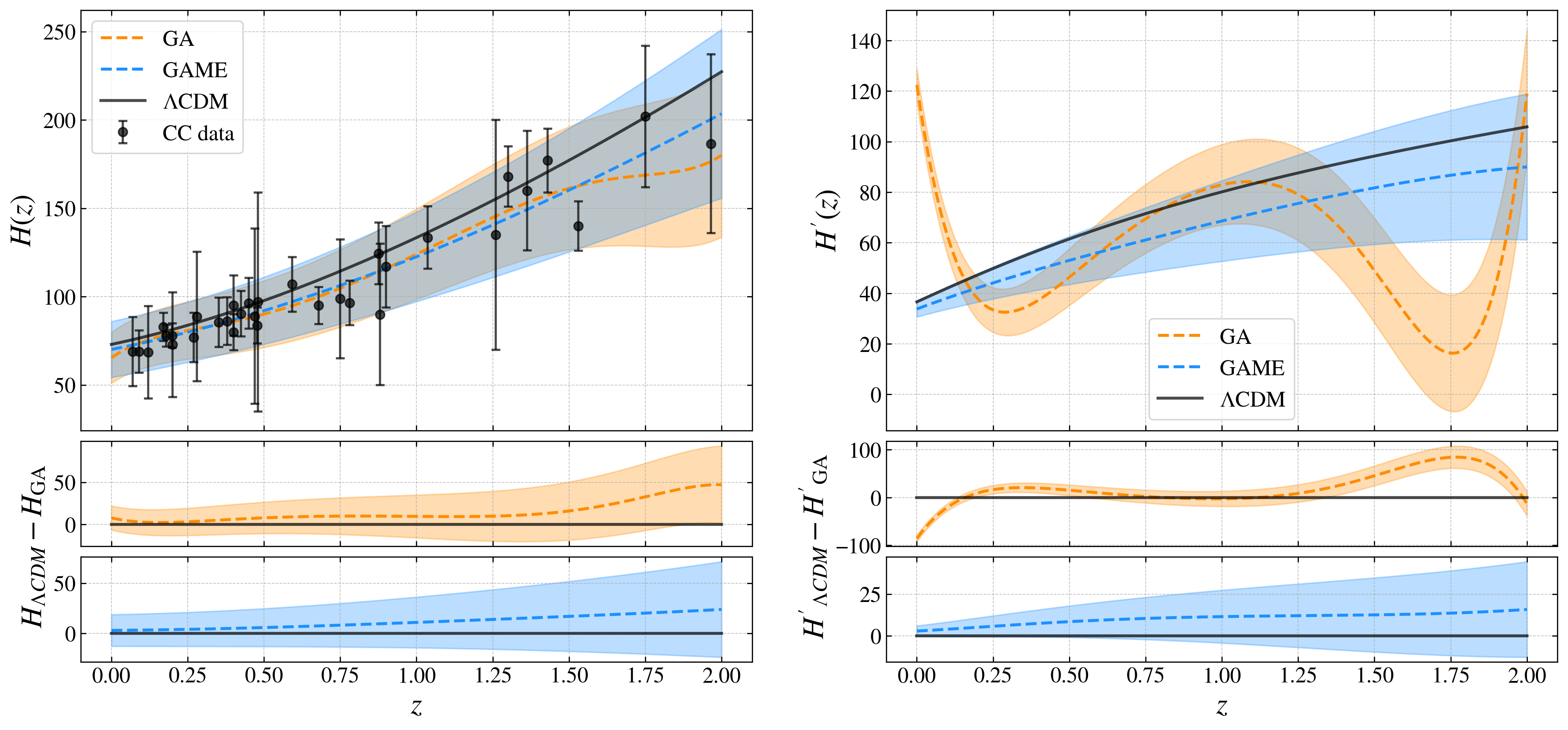}
\caption{Comparison of $H(z)$ reconstruction and derivative estimation between the baseline GA and \texttt{GAME} methodologies using CC data. \textit{Left:} Reconstruction of the Hubble parameter $H(z)$. The solid black line indicates the flat $\Lambda$CDM model. The orange dashed line shows the baseline best-fit GA result, while the blue dashed line represents the \texttt{GAME} marginalised-ensemble reconstruction. Shaded regions between dotted lines indicate $1\sigma$ confidence intervals: a purely statistical confidence interval ($\delta f_{\rm PI}$) for the baseline GA, and a hybrid band combining the statistical uncertainty with the configuration ensemble variance ($\sigma_{\rm ens}$) for \texttt{GAME} (Section~\ref{ssec:totalerror}). \textit{Right:} The corresponding reconstruction of the derivative $H'(z)$. While the function reconstructions (left) are largely consistent, the derivative reconstruction highlights the enhanced stability of the \texttt{GAME} method (blue), which provides a smoother evolution compared to the baseline GA (orange) that exhibits more pronounced fluctuations, particularly at the boundaries of the redshift domain.}
\label{fig:real_H}
\end{figure}

Once the GA has converged, we identify the global minimum $\chi^2$ best-fit solution, $H_{\rm GA}(z)$, and then calculate $H_{\rm GAME}$ using the \texttt{GAME} weighted-averaged approach described in Section \ref{ssec:game}. Figure \ref{fig:real_H} presents a comparison of the reconstructed $H(z)$ and $H'(z)$ obtained from the baseline minimum best-fit function selection and from the \texttt{GAME} approach.

Both methods provide a good reconstruction of $H(z)$, although \texttt{GAME} appears slightly more stable at higher redshifts. For the derivative $H'(z)$, \texttt{GAME} derivation is significantly more consistent with $\Lambda$CDM than the baseline GA result. The observed behaviour is consistent with what one would expect from the structure of the two methodologies. The baseline GA selects the configuration that minimises the $\chi^2$ between the data and $H_{\rm GA}(z)$, and is, by definition, insensitive to the smoothness of the reconstructed function. \texttt{GAME}, on the contrary, weights configurations by the regularised estimator of Equation~\eqref{eq:estimator}, where $R_j$ penalises reconstructions that are too noisy. The weighted-average reconstruction $H_{\rm GAME}(z)$ remains nearly indistinguishable from the baseline $H_{\rm GA}(z)$, but its derivative is substantially smoother, revealing the contribution of the roughness penalty in those regions where the data alone do not constrain the shape of the function. Because GA is primarily driven by the $\chi^2$ statistic, we also observed that excluding data points with larger uncertainty bands $\sigma_i$ substantially enhances the reconstruction quality, in particular its constraining ability and the behaviour of $\delta f_{\rm PI}$ (Appendix \ref{app:lesspoints}).

This confirms that the reconstruction of $H(z)$ and its derivative is the appropriate framework for applying the \texttt{GAME} methodology. The dark energy equation of state $w(z)$ enters the cosmological background equations through the derivative $H'(z)$, as shown in Equation~\eqref{eq:de_eos}. Any feature of $w(z)$ that one might hope to use as a diagnostic of new physics is only as stable as the derivative of the reconstructed $H(z)$.

\subsubsection{Derivation of $w(z)$}\label{ssec:w_real}
Using Equation \eqref{eq:de_eos}, we derive the evolution of the dark energy equation of state $w(z)$ under the assumptions of GR and a flat FLRW metric. This derivation is particularly challenging because it requires not only the function $H(z)$ but also its first derivative $H'(z)$, making it highly sensitive to the smoothness of the reconstruction. We utilise the $H_{\rm GAME}$ and $H'_{\rm GAME}$ profiles obtained from the marginalised ensemble of CC data to compute $w(z)$ assuming the two priors on $\Omega_{m,0}$ listed in \ref{list:Omegampriors}. The uncertainties on $w(z)$ are computed via standard error propagation, combining the variance from \texttt{GAME} reconstruction ($\sigma_H$) with the reported uncertainties on $\Omega_{m,0}$.
\begin{figure}[h]
\centering
\includegraphics[width=12cm]{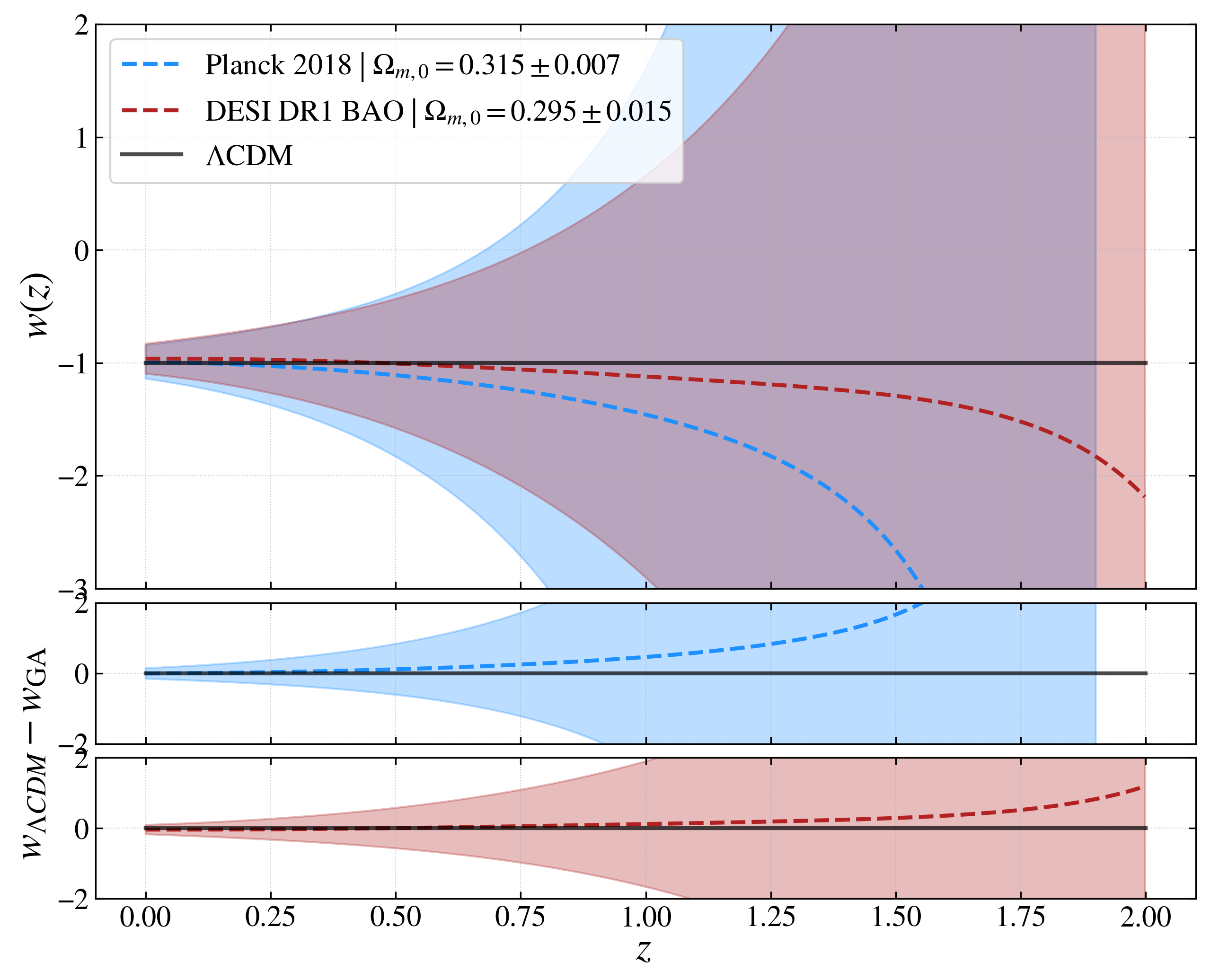}
\caption{Optimised estimation of the dark energy equation of state $w(z)$ derived from the \texttt{GAME}-reconstructed $H(z)$ using CC data. The derivation assumes a flat FLRW metric and is performed for two different priors on the matter density parameter $\Omega_{m,0}$. The solid black line indicates the standard $\Lambda$CDM value ($w = -1$). The blue dashed line represents the derivation assuming the Planck 2018 CMB prior, while the red dashed line assumes the DESI DR1 BAO prior. Shaded regions indicate the effective $1\sigma$ uncertainty bands obtained via error propagation of the total hybrid \texttt{GAME} uncertainty (Section~\ref{ssec:totalerror}) on $H(z)$ through Equation~\eqref{eq:de_eos}. At low redshifts ($z \lesssim 1$), both derivations produce tight constraints compatible with the cosmological constant. At higher redshifts, the reconstruction becomes highly sensitive to the assumed matter density and the sparsity of the data, leading to a significant widening of the uncertainty bands and a divergence, though remaining statistically consistent with $\Lambda$CDM.}
\label{fig:real_w}
\end{figure}

Figure \ref{fig:real_w} shows the estimated evolution of $w(z)$. At low redshifts ($z \lesssim 1$), where dark energy dominates the total energy budget, the reconstruction is robust and tightly constrained. Regardless of the assumed prior on $\Omega_{m,0}$, the results are in excellent agreement with the standard $\Lambda$CDM expectation ($w = -1$). At higher redshifts ($z \gtrsim 1.5$), however, the confidence regions begin to diverge significantly, indicating a loss of constraining power. This behavior is intrinsic to non-parametric derivations of $w(z)$ in a spatially flat universe. In such frameworks, the equation of state is determined by the logarithmic derivative of the dark energy density, $d\ln\Omega_{\rm de}/dz \propto \Omega'_{\rm de}/\Omega_{\rm de}$. As we approach the matter-dominated era ($z \gtrsim 1.5$), $\Omega_{\rm de}(z)$ is obtained as the difference between the total expansion rate and the matter contribution: $\Omega_{\rm de}(z) \approx H(z)^2/H_0^2 - \Omega_{m,0}(1+z)^3$. In this regime, that difference tends to zero ($\Omega_{\rm de} \to 0$), causing the logarithm in the expression for $w(z)$ toward infinity. As a result, the large broadening of the confidence intervals at high redshift is fundamentally caused by how uncertainties propagate near this mathematical singularity. This propagation strongly amplifies the measurement noise originating from sparser high-redshift observational data, which are themselves poorly constrained in that regime, generating apparent singularities even for small deviations in $H(z)$ or $\Omega_{m,0}$. 

Despite the loss of constraining power at high $z$, \texttt{GAME} methodology produces a robust estimate of the present-day equation of state. Adopting the DESI prior on $\Omega_{m,0}$, which favors a lower matter density, we obtain our tightest constraint:
\begin{equation}\label{eq:w0real}
    w(0) = -0.963 \pm 0.133.
\end{equation}
This measurement confirms that, with current CC data, the local universe remains fully compatible with a cosmological constant. The apparent deviation from $\Lambda$CDM observed in the reconstruction at high $z$ is not statistically meaningful given the large uncertainties, but it highlights the critical importance of precise $H(z)$ measurements and accurate local matter density priors for extending dark energy tests into the matter dominated era.

\subsection{Testing $\Lambda$CDM with Type Ia Supernovae}
As an additional independent real-data probe, we now apply the \texttt{GAME} methodology to a different cosmological observable: the distance modulus $\mu(z)$ of SNe Ia. The dataset we have used comes from the joint Pantheon+ and SH0ES data release \cite{Brout:2022, Riess:2021, Scolnic:2021}. Specifically, we utilise the standardised SNe Ia distance moduli and the full statistical and systematic covariance matrices. This test shows that \texttt{GAME} is not tailored to a single dataset, but can be applied to probes with very different sampling and noise properties: CC are sparse, heterogeneous direct measurements of $H(z)$, while the SNe sample is dense, highly correlated, and has comparatively low noise. Therefore, reconstructing the distance modulus is an additional test for the algorithm's ability to model an integrated observable while simultaneously managing a full non-diagonal covariance matrix, differently to what done with CC. We should stress that it is important to test our pipeline with realistic data sets, which by their nature are noisy, as the noise can in fact affect the performance and robustness of symbolic regression code \cite{matsubara2022rethinking}.

\subsubsection{Reconstruction of $\mu(z)$}\label{ssec:mu}
\begin{figure}[h]
\centering
\includegraphics[width=\linewidth]{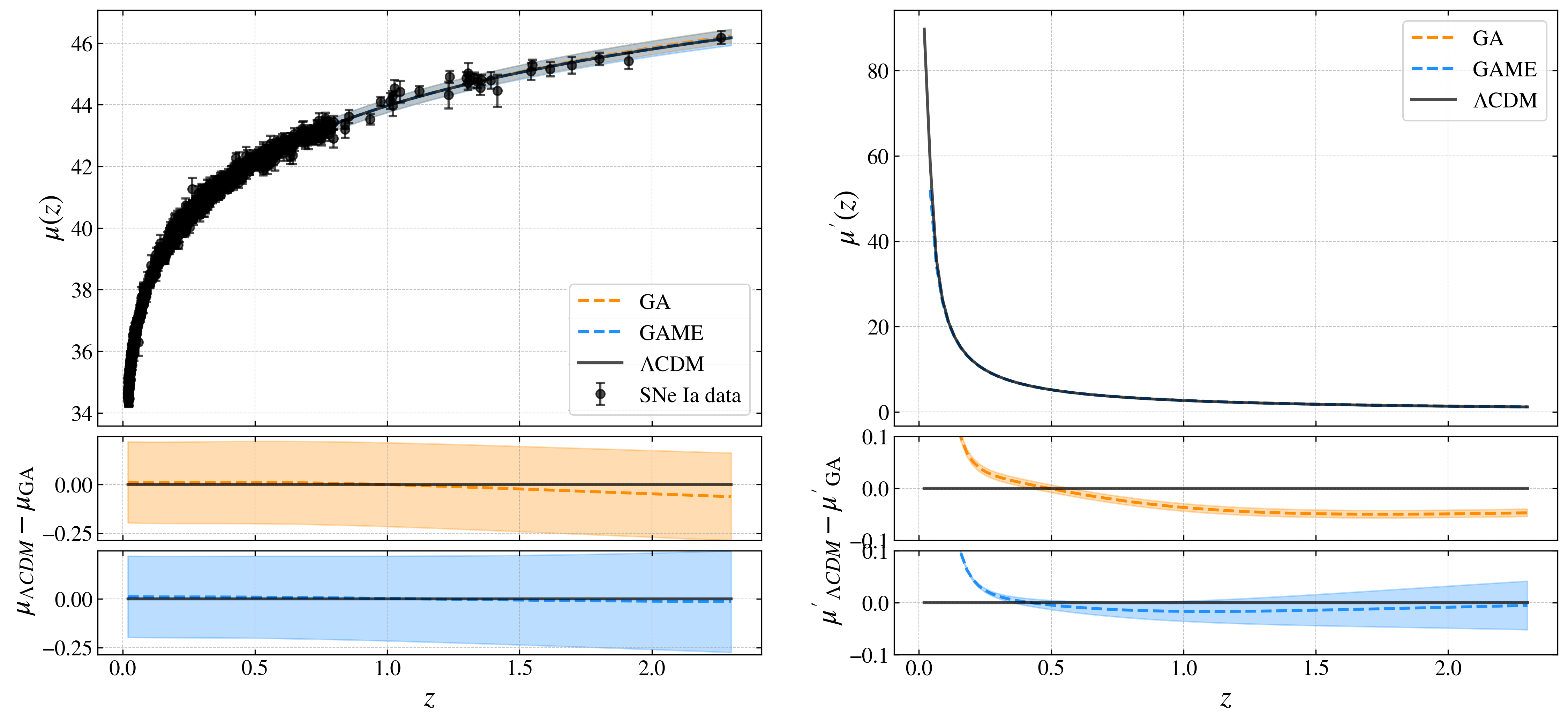}
\caption{Comparison of $\mu(z)$ reconstruction and derivative estimation between the baseline GA and \texttt{GAME} methodologies using SNe Ia data. \textit{Left:} Reconstruction of the distance modulus $\mu(z)$. The solid black line indicates the flat $\Lambda\text{CDM}$ model. The orange dashed line shows the baseline best-fit GA result, while the blue dashed line represents the GAME marginalised-ensemble reconstruction. Shaded regions between dotted lines indicate $1\sigma$ confidence intervals: a purely statistical confidence interval ($\delta f_{\rm PI}$) for the baseline GA, and a hybrid band combining the statistical uncertainty with the configuration ensemble variance ($\sigma_{\rm ens}$) for \texttt{GAME} (Section~\ref{ssec:totalerror}). \textit{Right:} The corresponding reconstruction of the derivative $\mu'(z)$. Both reconstructions correctly reproduce the steep growth as $z\to0$ driven by the asymtptotic behaviour (Equation~\eqref{eq:muprime}); for this reason the residuals are unstable at very low redshift, where even infinitesimally small fractional errors are amplified by the $1/z$ factor. From $z\sim0.25$ onward the residuals stabilise and the smoother evolution provided by \texttt{GAME} (blue) relative to the baseline GA (orange) becomes apparent.}
\label{fig:real_mu}
\end{figure}

The dataset we have used provides $N_{\rm SNe} = 1701$ light curve distance moduli over the redshift range $0.001\lesssim z\lesssim 2.3$. In particular, we drop the SNe Ia in Cepheid-host galaxies (used by SH0ES to anchor the distance ladder). In addition, following the SH0ES and Pantheon+ approach, we keep only SNe Ia at $z>0.023$, excluding sources affected by peculiar-velocity noise. After this filtering, the inclusion of the full Pantheon+SH0ES covariance matrix is straightforward via the correlated $\chi^2$ of Equation~\eqref{eq:correlatedchi2} in \texttt{GAME}. We fix the GA hyperparameter ensemble (Appendix \ref{app:ga_hyperparams}) and let the algorithm converge. We identify the global minimum $\chi^2$ best-fit solution $\mu_{\rm GA}$ and then calculate $\mu_{\rm GAME}$ similarly to what we did for $H(z)$. Figure~\ref{fig:real_mu} shows the reconstructed $\mu(z)$ and $\mu'(z)$ together with the Pantheon+SH0ES data points and the expected curve from $\Lambda$CDM.
As observed for $H(z)$, both methods provide a good reconstruction of $\mu(z)$, and even in this case \texttt{GAME} appears to be slightly more stable at higher redshift. For the derivative $\mu'(z)$ though, both reconstructions appear to blow up at very low redshifts ($z\ll1$) when looking at their residuals. This happens because, at very low redshifts, the luminosity distance is dominated by the linear Hubble flow ($D_L\propto z$). Because the distance modulus is logarithmic ($\mu\propto \log_{10}{D_L}\propto\log_{10}{z}$, see Equation~\eqref{eq:m_B}), its derivative contains a singularity at $z=0$:
\begin{equation}\label{eq:muprime}
    \mu'(z)\approx\dfrac{5}{\ln{10}}\dfrac{1}{z}.
\end{equation}
Both GA and \texttt{GAME} reconstructions are correctly reproducing this $1/z$ asymptotic behaviour, but because of the $1/z$ term, even an infinitesimally small fractional error is amplified to infinity as $z\to 0$. Since we are not imposing any prior on the $\mu(z)$ reconstruction, and given that the catalogue has been filtered from $z\gtrsim 0.02$, it becomes complicated for a symbolic regression algorithm to grasp both the data fit and the correct growth rate with redshift after only a few data points. Either way, from $z\sim 0.25$ onward it is possible to observe how also the residuals of $\mu'(z)$ stabilise, and the improvement given by \texttt{GAME} becomes especially clear. The reasons for this stabilisation is the same as in Section~\ref{ssec:H_real}: \texttt{GAME} weights configurations by a roughness penalty that takes into account the smoothness of the reconstructed function and suppresses noisy behaviour in the derivative. This stability is essential for the derivation of $H(z)$ from SNe Ia.

\subsubsection{Derivation of $H(z)$ and $w(z)$}
The distance modulus $\mu(z)$ is related to the luminosity distance $D_L(z)$ through:
\begin{equation}
    \mu(z) = 5\log_{10}\left({\dfrac{D_L(z)}{\rm{Mpc}}}\right)+25\to \dfrac{D_L(z)}{\rm Mpc} = 10^{\frac{\mu(z)}{5}-5}.
\end{equation}
Assuming a flat FLRW universe, the luminosity distance is related to the comoving distance $D_C(z)$ by:
\begin{equation}\label{eq:com_lum_distance}
    D_L(z) = (1+z)D_C(z).
\end{equation}
And the comoving distance itself is defined as the integral of the inverse Hubble rate:
\begin{equation}\label{eq:comoving_distance}
    D_C(z) = c\int_0^z\dfrac{dz'}{H(z')}.
\end{equation}
Differentiating Equation~\eqref{eq:comoving_distance} with respect to $z$ gives $dD_C/dz=c/H(z)$. Differentiating Equation~\eqref{eq:com_lum_distance} and substituting, then solving for $H(z)$, we obtain an expression where we can plug our reconstructed $\mu_{\rm GAME}(z)$ and $\mu'_{\rm GAME}(z)$:
\begin{equation}\label{eq:Hfrommu}
    H(z) = \dfrac{\dfrac{c(1+z)}{D_L(z)}}{\dfrac{\ln{10}}{5}\mu'(z)-\dfrac{1}{1+z}}.
\end{equation}
From this expression, we utilise our functions obtained from the marginalised ensemble of SNe Ia data and we compute $H(z)$ with associated uncertainty computed via standard error propagation using the variance from \texttt{GAME} reconstruction ($\sigma_\mu$).
\begin{figure}[h]
\centering
\includegraphics[width=12cm]{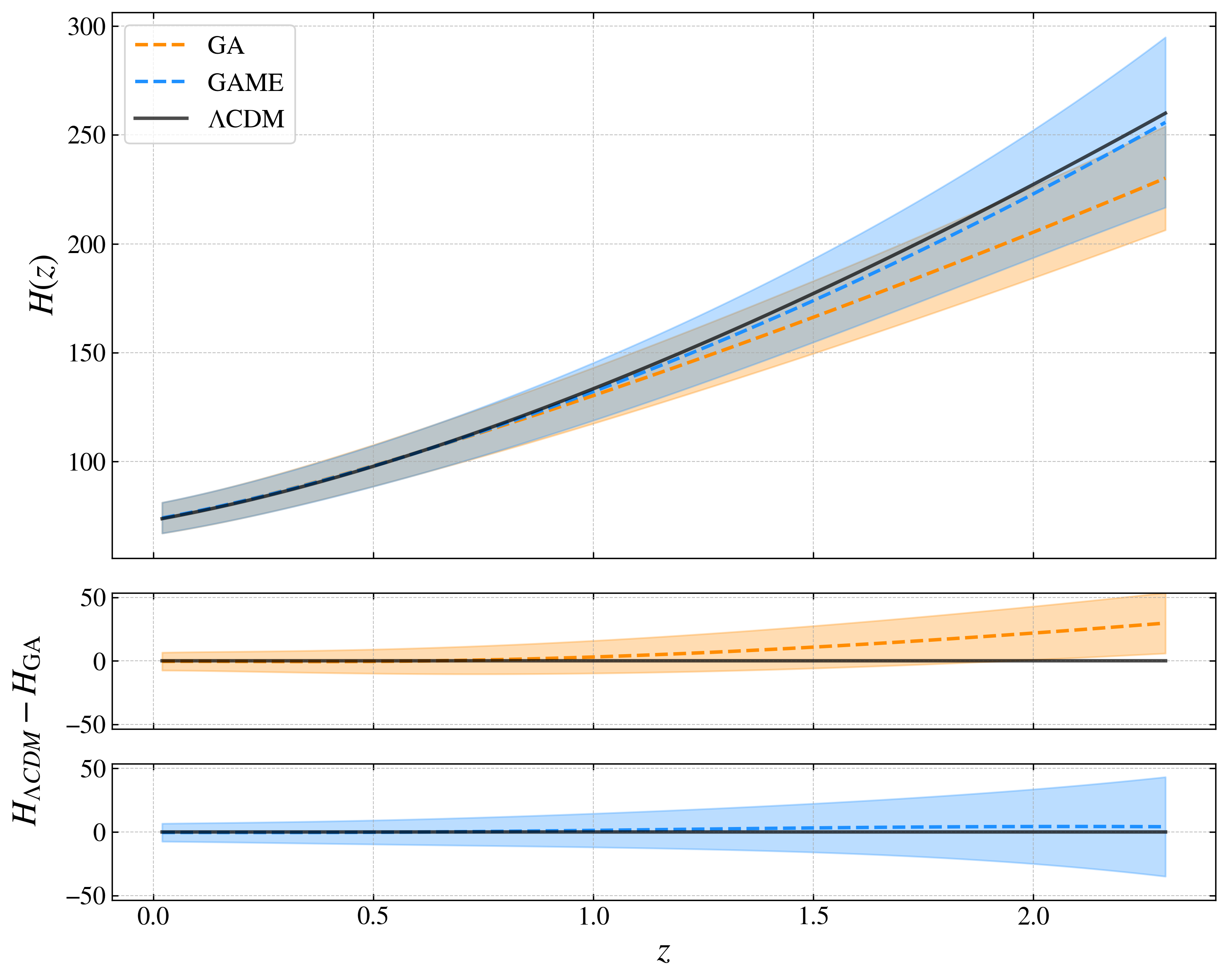}
\caption{Hubble rate $H(z)$ derived by differentiating the reconstructed $\mu(z)$ of Figure~\ref{fig:real_mu} through Equation~\ref{eq:Hfrommu}. The solid black line indicates the flat $\Lambda$CDM expectation. The orange dashed line shows the $H(z)$ from the baseline GA reconstruction of $\mu(z)$, while the blue dashed line shows the $H(z)$ derived from the \texttt{GAME} reconstruction. Shaded regions indicate the effective $1\sigma$ uncertainty bands obtained via error propagation of the total hybrid \texttt{GAME} uncertainty (Section~\ref{ssec:totalerror}) on $\mu(z)$ through Equation~\eqref{eq:Hfrommu}. The reconstruction recovers the fiducial expansion history over the full redshift range, and the \texttt{GAME} approach delivers a tighter and more stable band than the baseline GA at high redshift, where the differentiation step amplifies the uncertainty.}
\label{fig:real_H_from_mu}
\end{figure}
Figure~\ref{fig:real_H_from_mu} shows the estimate evolution of $H(z)$ using SNe Ia data. Compared with the direct CC reconstruction of Figure~\ref{fig:real_H}, the SNe Ia-derived $H(z)$ is more tightly constrained at low and intermediate redshift, despite being a derived rather than a directly measured quantity. This shows how utilising a different dataset, with denser and more precise measurements, can help \texttt{GAME} provide a more robust reconstruction and computation of derived quantities. We also show the comparison with the single best-fit solution from baseline GA, and it shows how even at high redshifts (higher than the boundaries considered for CC data) \texttt{GAME} maintains much more stability.
Finally, we can plug the $H(z)$ from SNe Ia and also estimate the evolution of $w(z)$. This allows us to assess the algorithm's performance through a full $\mu(z)\to H(z)\to w(z)$ chain, which involves two derivatives of the originally fitted observable. This derivation is particularly demanding because $w(z)$ depends on $H'(z)$, which in turn requires $\mu'(z)$ and $\mu''(z)$, making the result especially sensitive to the smoothness of the original $\mu(z)$ reconstruction. In order to be consistent with the dataset used, besides the Planck and DESI priors listed in \ref{list:Omegampriors}, we also include the Pantheon+ on $\Omega_{m,0}$:
\begin{itemize}
    \item \textbf{Pantheon+} \cite{Brout:2022}: $\Omega_{m,0} = 0.334 \pm 0.018$, representing the SNe Ia constraint.
\end{itemize}
\begin{figure}[h]
\centering
\includegraphics[width=12cm]{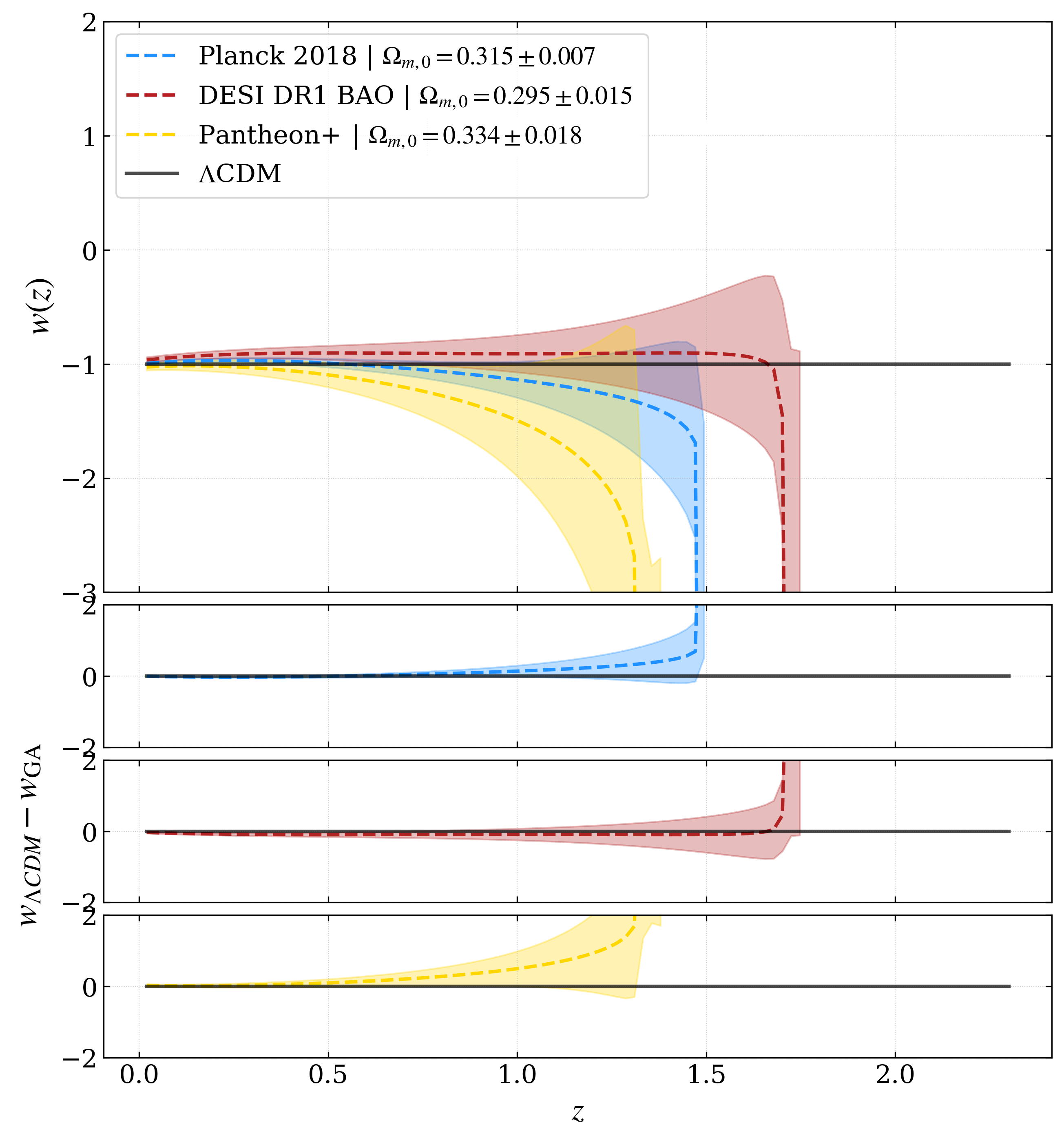}
\caption{Dark energy equation of state $w(z)$ derived from the reconstructed $\mu(z)$ from \texttt{GAME} using SNe Ia data, through the double derivative process $\mu(z)\to H(z)\to w(z)$. The derivation assumes a flat FLRW metric and GR, and is performed for three different priors on the matter density parameter $\Omega_{m,0}$. Shaded regions indicate the effective $1\sigma$ uncertainty bands obtained via error propagation of the total hybrid \texttt{GAME} uncertainty (Section~\ref{ssec:totalerror}) on $H(z)$ through Equation~\eqref{eq:de_eos}. At low redshifts ($z\lesssim0.7$) the three derivations produce tight constraints compatible with the cosmological constant, with the dense low $z$ SNe Ia sample producing a noticeably tighter band than the corresponding CC-derived reconstruction in Figure~\ref{fig:real_w}. At higher redshifts the confidence regions broaden and the reconstruction diverges: this reflects the structural singularity at $\Omega_{de}\to0$ inherent to the non-parametric derivation of $w(z)$ in a flat FLRW, here amplified by the fact that $w$ is obtained through a second derivative of the underlying reconstructed observable.}
\label{fig:real_w_from_mu}
\end{figure}
Figure~\ref{fig:real_w_from_mu} shows the estimate evolution of $w(z)$, this time coming out of the SNe Ia dataset. It is possible to observe several analogies with the reconstruction obtained from CC. At low redshifts ($z \lesssim 0.7$) the reconstruction is robust and even more tightly constrained (thanks to the fact that SNe Ia is a more dense and precise dataset if compared to CC, and as seen in Figure~\ref{fig:real_H_from_mu} the uncertainty band are tighter than its CC counterpart. The results are in excellent agreement with the standard $\Lambda$CDM expectation. At higher redshifts ($z\gtrsim 1$), however, we start observing the same divergence of both the mean and uncertainty regions. Once again, this behaviour is intrinsic to non-parametric derivations of $w(z)$ in a spatially flat universe, because of the logarithm in the expression for $w(z)$ going toward infinity.
Even in this case, despite the loss of constraining power and the divergence of the reconstruction at high $z$, \texttt{GAME} methodology produces a robust estimate of the dark energy equation of state through a second derivative expression. In this case, the tightest constraint is obtained by adopting the Planck prior on $\Omega_{m,0}$, at a redshift of $z=0.02$ remembering that, as explained in Section~\ref{ssec:mu}, we exclude sources with $z<0.023$ to remove peculiar velocity-dominated SNe Ia, so $w(0.02)$, strictly, evaluated on the smooth extrapolation of the reconstruction toward $z=0$:
\begin{equation}
    w(0.02) = -0.995 \pm 0.024.
\end{equation}
This measurement highlights that with current SNe Ia data, the local universe remains fully compatible with a cosmological constant. Moreover, using \texttt{GAME} in the right conditions and with a detailed dataset allows for stable derivations even when involving the second derivative of the reconstructed functions.

\subsection{Validation with controlled mock dataset}
To further validate the \texttt{GAME} methodology, we test its robustness using mock data. We do not work with a specific future survey in mind, but we rather focus on obtaining results having under control the quality of the data, as well as having a known fiducial cosmology, which needs to be recovered by our approach.

Our mock data are generated as follows:
\begin{enumerate}
    \item we assume a standard fiducial cosmological model ($\Lambda$CDM), which provides the ground truth for our $H^{\rm fid}(z)$;
    \item we assume to observe $N_{\rm data}=15$ data points distributed between redshift $0$ to $2$;
    \item for each observation, we assume a Gaussian error $\sigma_H(z) = 2\% (1+z)\,H(z)$, with the redshift dependence capturing the typical loss of precision at higher redshifts found in spectroscopic surveys \cite{EuclidForecasts:2020}. The choice of the $2\%$ error factor wants to mimic an improvement on the precision of CC measurements from future observations, although we do not attempt to mimic any planned survey. More realistic forecasts of the results that will be achieved with future surveys are left to a follow-up work;
    \item at each $z_i$ where we assume our observations are performed, we obtain the observed value $H(z_i)$ from a normal distribution $\mathcal{N}(H^{\rm fid}(z),\sigma_H(z_i))$.
\end{enumerate}

We examined both uniformly spaced and randomly sampled redshift distributions and obtained comparable results; in what follows, we show the results for the uniform case.

Other than assessing the performance of our method having under control the quality of the data, working with mock data also allows us to test the accuracy of \texttt{GAME} that could be reached with future surveys. Indeed, knowing the fiducial cosmology and, therefore, the expected final result, we can test if \texttt{GAME} can accurately recover the underlying function and its derivative from noisy synthetic data, without generating artificial features, thus confirming the method's reliability for applications in the precision cosmology era.

\subsubsection{Reconstruction of $H(z)$}\label{ssec:H_mock}
\begin{figure}[h]
\centering
\includegraphics[width=\linewidth]{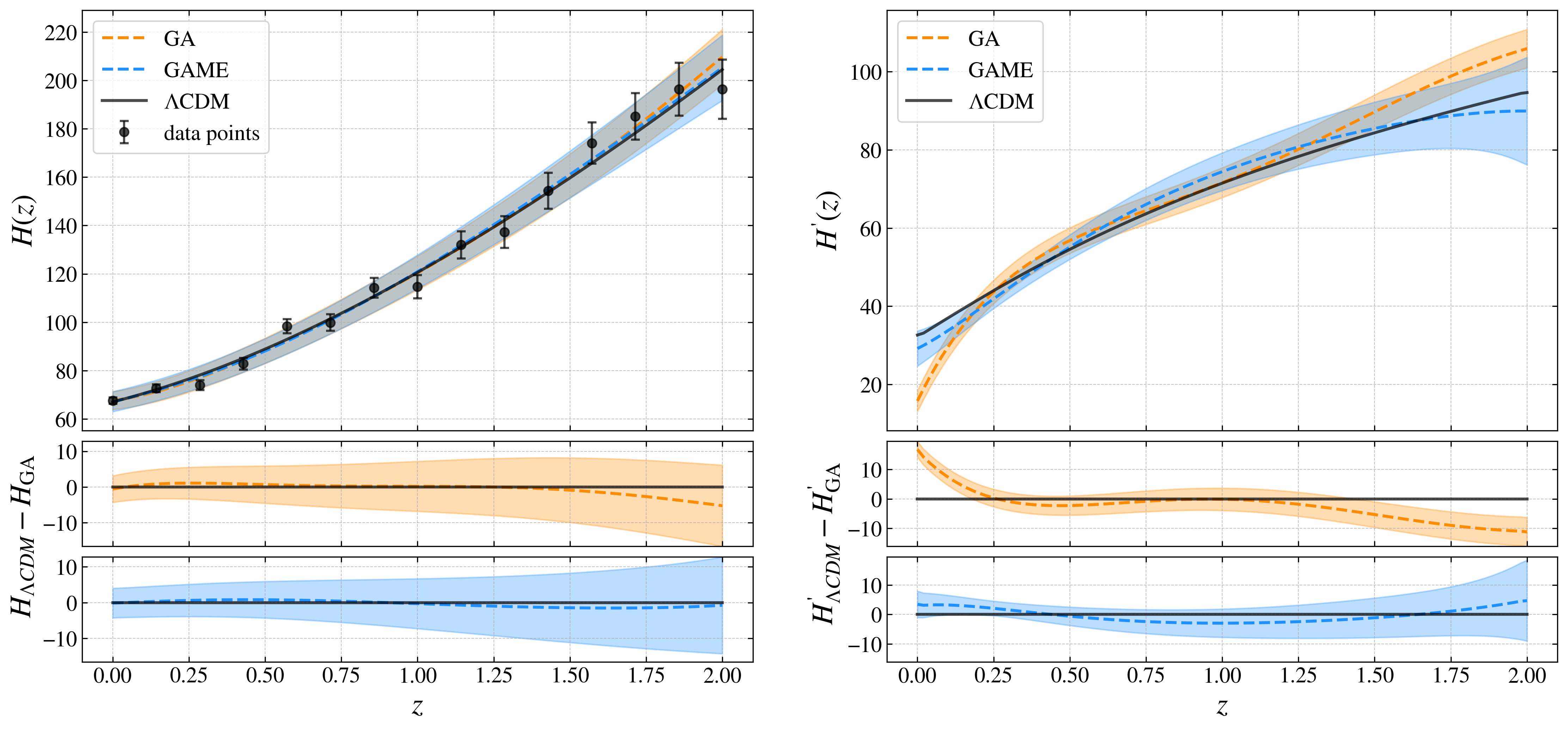}
\caption{Comparison of $H(z)$ reconstruction and derivative estimation between the baseline GA and \texttt{GAME} methodologies using a mock dataset. \textit{Left:} Reconstruction of the Hubble parameter $H(z)$ from 15 mock data points generated on a flat $\Lambda$CDM fiducial (solid black line). The points include a measurement uncertainty of $2\%(1+z)$. The orange dashed line shows the baseline best-fit GA result, while the blue dashed line represents the \texttt{GAME} averaged reconstruction. Shaded regions indicate $1\sigma$ confidence intervals: a purely statistical confidence interval ($\delta f_{\rm PI}$) for the baseline GA, and a hybrid band combining the statistical uncertainty with the configuration ensemble variance ($\sigma_{\rm ens}$) for \texttt{GAME} (Section~\ref{ssec:totalerror}). \textit{Right:} The corresponding estimate of the derivative $H'(z)$. As observed with real data (Figure \ref{fig:real_H}), \texttt{GAME} reconstruction maintains the smoothness of the derivative $H'(z)$ of the fiducial $\Lambda$CDM input, against which the reconstruction is being benchmarked, successfully recovering the $\Lambda$CDM fiducial without the high-redshift oscillations present in the single best-fit GA run.}
\label{fig:mock_H}
\end{figure}

Figure \ref{fig:mock_H} compares the reconstruction of $H(z)$ and its derivative $H'(z)$ using the baseline GA best-fit versus the \texttt{GAME} averaged function. Both methods successfully reconstruct the Hubble parameter $H(z)$, closely following the fiducial $\Lambda$CDM line. Once again, the strength of the \texttt{GAME} methodology becomes apparent in the derivative. Symbolic regression techniques can be sensitive to local noise: such noise may induce spurious local structure in the reconstructed function $H(z)$, which can lead to unstable, strongly oscillating derivatives $H'(z)$ even when $H(z)$ itself looks smooth. The \texttt{GAME} reconstruction effectively suppresses these oscillations, reproducing the fiducial derivative more faithfully than the single GA best-fit, particularly at the boundaries ($z \approx 0$ and $z \approx 2$).

\begin{figure}[h]
\centering
\includegraphics[width=\linewidth]{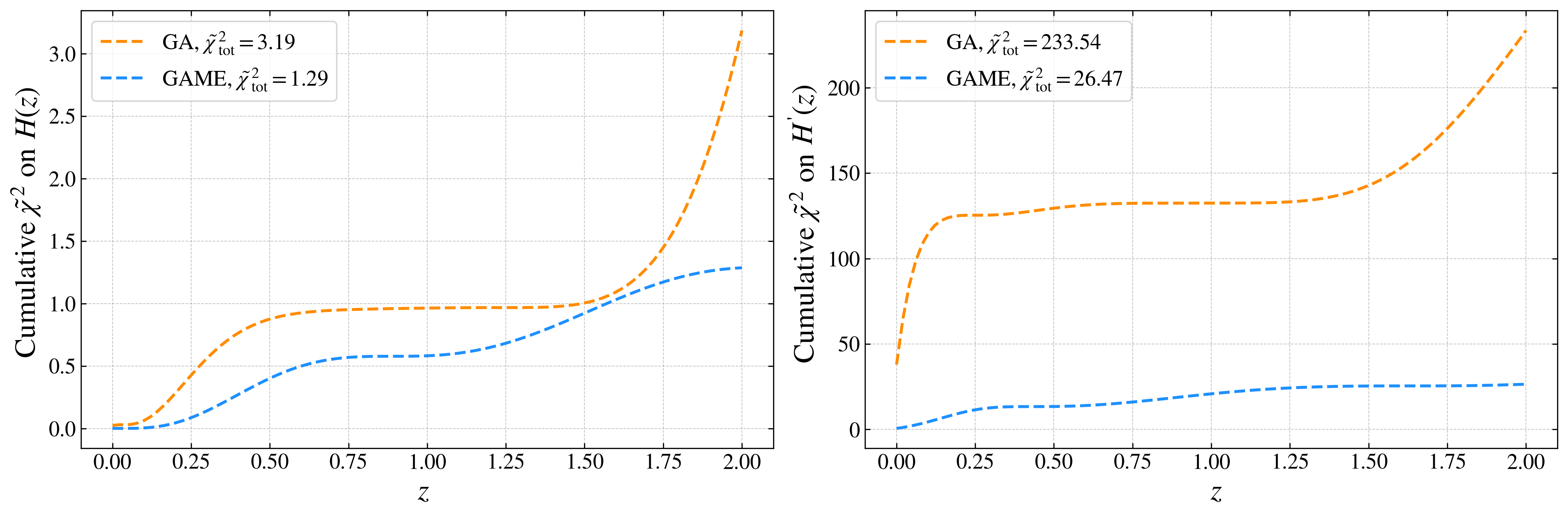}
\caption{Cumulative performance of the reconstruction methodologies against the assumed fiducial model ($\Lambda$CDM). The panels show the cumulative $\tilde\chi^2_{\rm tot}$ as a function of redshift $z$, calculated between the reconstructed functions and the underlying target model. \textit{Left:} Cumulative $\tilde\chi^2_{\rm tot}$ for the Hubble rate $H(z)$. The blue line (\texttt{GAME} approach) reaches a final value of $\tilde\chi^2_{\rm tot}=1.29$, representing a $\sim 60\%$ improvement over the orange line (baseline GA best-fit, $\tilde\chi^2_{\rm tot}=3.19$). \textit{Right:} Cumulative $\tilde\chi^2_{\rm tot}$ for the derivative $H'(z)$. Here the advantage of the proposed method is more clear: the baseline GA accumulates large errors due to instability ($\tilde\chi^2_{\rm tot}= 233.54$), while \texttt{GAME} maintains robustness ($\tilde\chi^2_{\rm tot}=26.47$), resulting in a $\sim 89\%$ improvement in accuracy.}
\label{fig:cumulative_chi2_mock}
\end{figure}

As for the test function $f(x)$ in Section~\ref{ssec:f_mock} we quantify this improvement in stability by computing the cumulative statistic $\tilde\chi^2_{\rm tot}$. In Figure~\ref{fig:cumulative_chi2_mock} we show the evolution of the cumulative $\tilde\chi^2_{\rm tot}$ for both $H(z)$ and $H'(z)$.
For $H(z)$ the improvement is visually subtle in the reconstruction plot, but the cumulative estimator reveals a quantitative gain of $\sim60\%$, with $\tilde\chi^2_{\rm tot}$ decreasing from $\tilde\chi^2_{\rm tot, GA} = 3.19$ to $\tilde\chi^2_{\rm tot, GAME}=1.29$. For the derivative $H'(z)$ the advantage of the \texttt{GAME} approach is much clearer, achieving an improvement of $\sim89\%$, as $\tilde\chi^2_{\rm tot}$ is reduced from $\tilde\chi^2_{\rm tot, GA} = 233.54$ to $\tilde\chi^2_{\rm tot, GA} = 26.47$. Besides representing a substantial improvement and notable increase in the reliability of recovering the underlying physical models, this refined stability is essential for deriving the equation of state $w(z)$, which depends directly on the gradient of the expansion history.

\subsubsection{Derivation of $w(z)$}
Using the \texttt{GAME}-reconstructed $H(z)$ and $H'(z)$ from the mock dataset, we compute the dark energy equation of state $w(z)$ by applying the same procedure described in Section \ref{ssec:w_real}. Since the mock observations are generated under a $\Lambda$CDM model with $w=-1$, this acts as a null test for our pipeline: we verify whether the algorithm introduces any systematic bias or signatures of dynamical dark energy.
\begin{figure}[h]
\centering
\includegraphics[width=12cm]{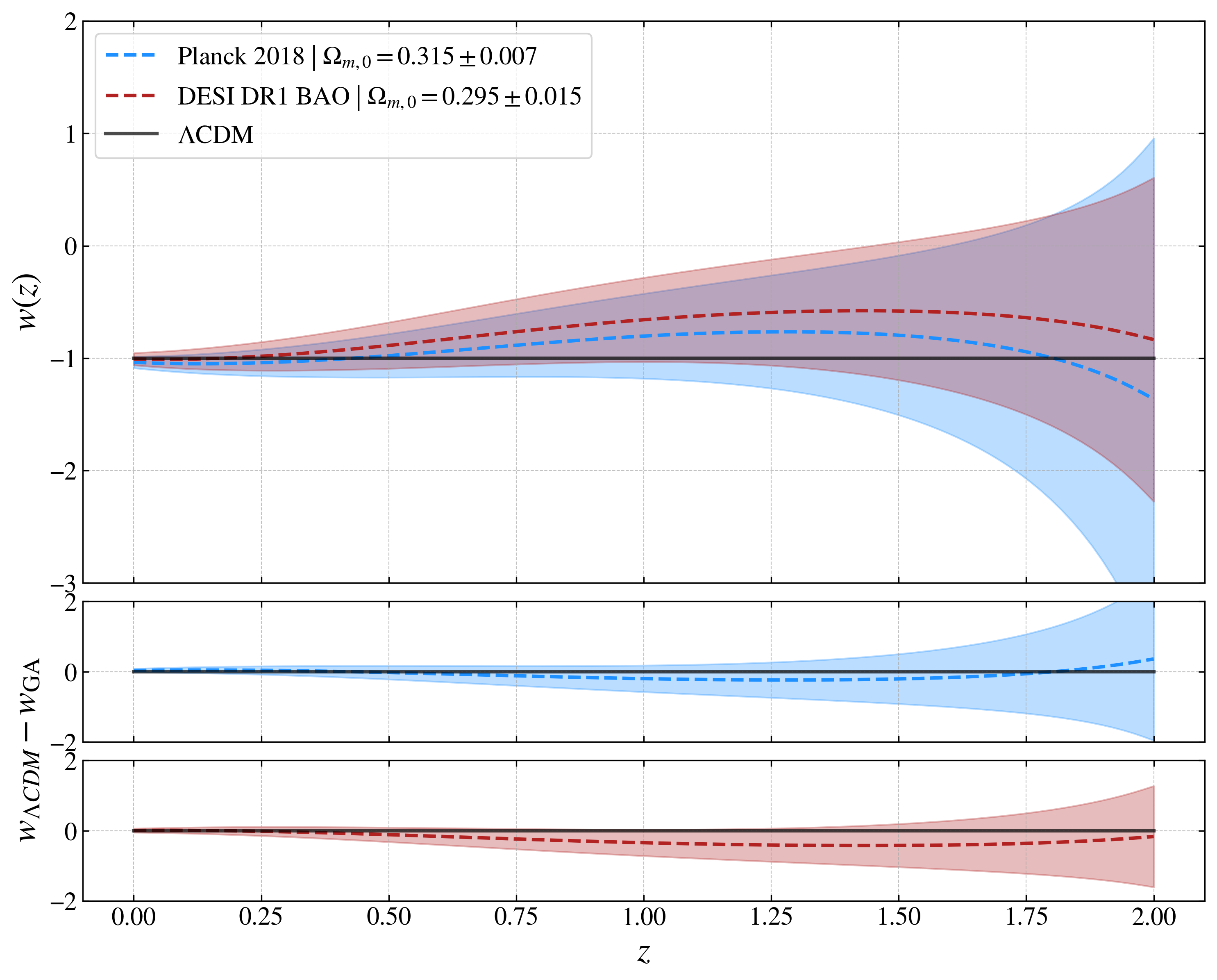}
\caption{Optimised reconstruction of the dark energy equation of state $w(z)$ derived from the \texttt{GAME}-reconstructed mock $H(z)$. The derivation assumes a flat FLRW metric and is performed for two different priors on the matter density parameter $\Omega_{m,0}$. The solid black line indicates the input fiducial $\Lambda$CDM model ($w = -1$). The blue dashed line represents the derivation assuming the Planck prior, while the red dashed line assumes the DESI DR1 BAO prior. Shaded regions indicate the effective $1\sigma$ uncertainty bands obtained via error propagation of the total hybrid \texttt{GAME} uncertainty (Section~\ref{ssec:totalerror}) on $H(z)$ through Equation~\eqref{eq:de_eos}. The reconstruction successfully recovers the fiducial value $w=-1$ within $1\sigma$ over the entire redshift range. Notably, the constraints at $z=0$ are significantly tighter than those obtained from current data, demonstrating the potential of \texttt{GAME} to leverage the precision of future surveys.}
\label{fig:mock_w}
\end{figure}
Figure \ref{fig:mock_w} shows the resulting $w(z)$ obtained adopting the same priors on $\Omega_{m,0}$ as those used for the derivation with real data. The reconstruction is fully consistent with the fiducial value $w=-1$ across the entire redshift range. The uncertainty bands naturally widen at higher redshifts due to the combined effects of the error propagation in Equation \eqref{eq:de_eos} and the decreasing relative importance of dark energy relative to matter. Most notably, the constraint at $z=0$ is significantly improved compared to what is obtainable with current CC measurements. Assuming Planck prior for $\Omega_{m,0}$, we obtain a precise estimate of the present-day equation of state:
\begin{equation}
    w(0) = -1.037 \pm 0.051.
\end{equation}
This result represents a factor of $\sim 2.6$ improvement in precision compared to the result obtained from current data in Section \ref{ssec:w_real}. This test confirms that \texttt{GAME} is not only capable of unbiased reconstruction, but demonstrates the strong potential to leverage the higher statistical power of datasets with improved precision to place tight, non-parametric constraints on the nature of dark energy, regardless of the prior on $\Omega_{m,0}$. Note that the SNe Ia constraint is already at this level of precision with current data, thanks to the much large sample size; the present analysis quantifies what an improvement of comparable magnitude on direct $H(z)$ measurements would deliver.

\section{Conclusions and outlook}
\label{sec:conclusions}
In this paper, we proposed \texttt{GAME}: an update to the baseline GA methodology for model-independent reconstructions of cosmological functions. While GA provide a versatile symbolic regression framework independent of specific parametrisations, a known weakness is the sensitivity of derived quantities (especially those involving derivatives) to stochasticity and hyperparameter choices. This sensitivity can lead to reconstructions that fit the data equally well but generate noticeably different derivatives, affecting consistency tests based on derived quantities like the dark energy equation of state $w(z)$.

To mitigate these issues, we introduced a model-averaging approach over an ensemble of GA best-fit reconstructions (\texttt{GAME}). We defined weights using a combined estimator $S_j=\chi^2_j+\lambda R_j$ (where $R_j$ is a roughness penalty) and selected the regularisation scale $\lambda$ via an L-curve method. The final reconstruction is the weighted average of all the analytical GA solutions associated to different hyperparameter configurations, improving stability and reducing overfitting in derivative-dependent observables. Consequently, we updated the uncertainty estimation to match the averaged framework by combining a measurement error component (estimated through a path-integral uncertainty estimation) with an ensemble error contribute (estimated through the weighted variance across configurations), constructing a conservative total uncertainty. 

This updated methodology also benefits from the significant increase in computational power available since the introduction of GA in cosmology. While early applications were often limited to $\mathcal{O}(10^3)$ generations and a single configuration due to CPU constraints, modern resources allow us to routinely exceed this by one or two orders of magnitude and by also exploring many configurations at the same time. This deeper exploration of the functional space, combined with our ensemble averaging, improves the convergence of the algorithm and reduces the risk of settling in local minima, a behaviour that is explicitly visualised by the convergence diagnostics across multiple configurations presented earlier in Figure~\ref{fig:many_iterations_chi2}.

We validated the method on a controlled example function $f(x)$, demonstrating that the marginalised ensemble reconstruction stabilises both the function and its derivative. Using a $\tilde\chi^2$ statistic against the true fiducial model, we quantified an improvement of $\sim 20\%$ in the reconstruction of the function itself and a substantial $\sim 95\%$ improvement in the accuracy of the derivative, highlighting the ability of \texttt{GAME} to reproduce underlying models. We then applied \texttt{GAME} to current data: directly reconstructing $H(z)$ from CC and $\mu(z)$ from SNe Ia data, and deriving in both cases $w(z)$ under FLRW and GR assumptions for different priors on $\Omega_{m,0}$. Within uncertainties, the reconstructed $w(z)$ is compatible with $\Lambda$CDM at low redshift. The loss of constraining power at higher redshift naturally widens the bands, highlighting the limitations of sparse $H(z)$ measurements and the importance of stable derivative reconstruction.

Finally, we tested \texttt{GAME} performance with controlled mock datasets validation. Even with a conservative error factor assumption, and while we keep using the same priors on $\Omega_{m,0}$, \texttt{GAME} recovers the fiducial $\Lambda$CDM behaviour with improved stability and produces a significantly tighter reconstruction of $w(z)$ at low redshift compared to present data. This illustrates the potential of GA reconstructions, when properly marginalised, to become a competitive tool for model-independent consistency tests with future, more precise surveys.

Several natural extensions of this work are possible. On the data side, a key next step is to incorporate the full covariance structure of CC (and other probes) directly into the GA likelihood, in order to assess and quantify the impact of data correlations. On the theoretical side, one can broaden the consistency analysis by relaxing assumptions such as spatial flatness and by jointly reconstructing multiple background and perturbation observables, enabling more direct tests of gravity and of the associated underlying theories. Taken together, \texttt{GAME} offers a practical and adaptable framework for stable, model-independent reconstructions of cosmological functions and their derivatives, making it a particularly promising tool for use in the precision era of upcoming large-scale surveys.

\acknowledgments
M.P. acknowledges support from the INFN project ``InDark''. M.P. also acknowledges the participation in the COST Action CA21136 ``Addressing observational tensions in cosmology with systematics and fundamental physics'' (CosmoVerse). M.M. acknowledges funding by the Agenzia Spaziale Italiana (\textsc{asi}) under agreement n. 2024-10-HH.0 and support from INFN/Euclid Sezione di Roma. S.N. acknowledges support from the research project PID2024-159420NB-C43 and the Spanish Research Agency (Agencia Estatal de Investigaci\'on) through the Grant IFT Centro de Excelencia Severo Ochoa No CEX2020-001007-S, funded by MCIN/AEI/10.13039/501100011033. M.P. also thanks A. Favale and F. De Luca for the useful discussions.

\section*{Code availability}
The implementation of the \texttt{GAME} methodology used in this work is publicly available at \url{https://github.com/matteoperonaci/GAME}; it is built on top of the Genetic Algorithms Python package of S. Nesseris (\url{https://github.com/snesseris/Genetic-Algorithms}), which implements the baseline GA of \cite{Bogdanos:2009} and is used for the comparisons reported throughout the paper. All scripts needed to reproduce the figures of this work are provided in the \texttt{GAME} repository.

\bibliographystyle{JHEP}
\bibliography{biblio}

\appendix

\section{Testing $\Lambda$CDM with a subset of current data}\label{app:lesspoints}
In the main analysis (Section \ref{ssec:H_real}), we adopted a conservative approach by utilising the full available dataset of CC up to $z \approx 2$, regardless of the precision of individual measurements. However, the CC dataset is heterogeneous, combining measurements from various surveys with different sensitivities and systematics. As GA is driven by the optimisation of the $\chi^2$ statistic, data points with large uncertainties contribute less to the fitness of the solution but can still introduce stochastic noise, particularly in the reconstruction of the derivative $H'(z)$. To verify the impact of these lower-precision measurements on our results, we performed an additional run of the \texttt{GAME} pipeline using a quality-cut subset of the data. In this subsample, we excluded data points with value $H > H_{\rm mean}+2\,\sigma_H$. This threshold removes the measurements with the lowest constraining power, which are often those most susceptible to large systematic effects or low signal-to-noise ratios.
\subsection{Impact on the reconstruction of $H(z)$}
The reconstruction of $H(z)$ and its derivative using this high-quality subset generates results that are fully consistent with the full-dataset analysis but differ enhancing more stability. While the reconstruction of $H(z)$ remains largely unchanged, confirming the robustness of \texttt{GAME} averaged reconstructions, the improvement is most notable in the derivative $H'(z)$. In the full analysis (Figure \ref{fig:real_H}), the derivative exhibits fluctuations at high redshifts ($z > 1.2$) where the data is sparser and noisier. By removing the points with large error bars, the resulting $H'(z)$ becomes significantly smoother and tracks $\Lambda$CDM prediction more closely. This suggests that the slight oscillatory features observed in the main analysis are likely artifacts driven by the scatter of low-precision data points rather than physical deviations from the standard model. Furthermore, the exclusion of these outliers decreases the path-integral uncertainty estimation $\delta f_{\rm PI}$.
\begin{figure}[h]
\centering
\includegraphics[width=\linewidth]{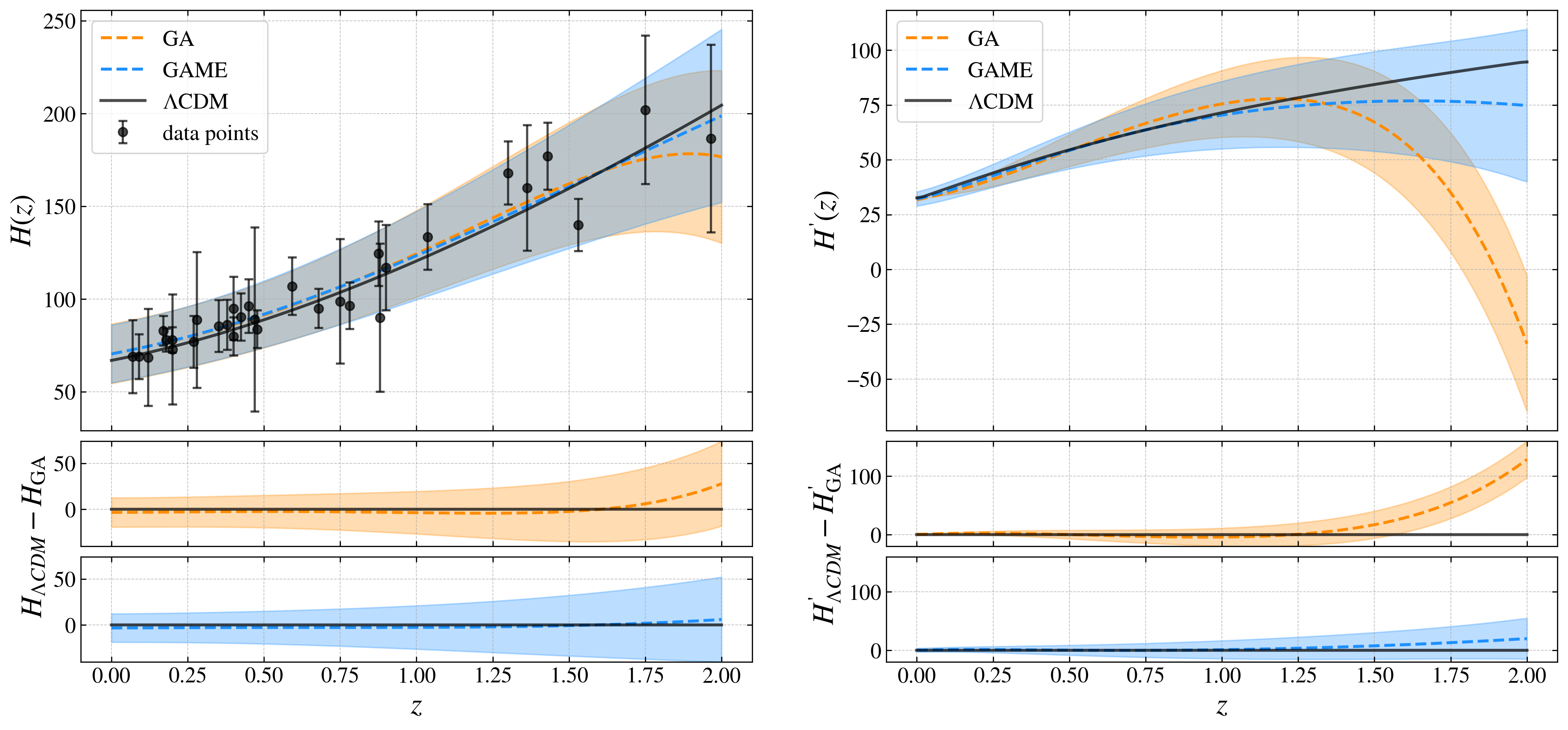}
\caption{Comparison of the reconstructed $H(z)$ and $H'(z)$ obtained from the quality-cut CC subset (excluding points with $H > H_{\rm mean}+2\,\sigma_H$). The plotting style matches Figure \ref{fig:real_H}: the solid black curve represents $\Lambda$CDM, the orange dashed curve is the baseline GA best-fit, and the blue dashed curve is the \texttt{GAME} averaged reconstruction. \textit{Left:} The $H(z)$ reconstruction remains stable and in excellent agreement with the result from the full dataset, demonstrating that the \texttt{GAME} approach is robust to the exclusion of low-quality measurements. \textit{Right:} The $H'(z)$ reconstruction becomes noticeably smoother relative to the full-sample case. Eliminating the high-uncertainty outliers mitigates the large fluctuations at high redshift seen in the main analysis, generating a derivative profile closer to the $\Lambda$CDM prediction and a lower path-integral uncertainty.}
\label{fig:less_H}
\end{figure}

\subsection{Impact on the derivation of $w(z)$}
The smoothness of $H'(z)$ has a direct consequence on the derivation of the dark energy equation of state $w(z)$. Since $w(z)$ depends on the ratio of terms involving both $H(z)$ and $H'(z)$, any unphysical fluctuation in the derivative is amplified in the equation of state. Using the subset, we observe that the divergence of the confidence intervals for $w(z)$ at $z > 1.5$ is slightly mitigated, although the mathematical singularity inherent to the vanishing dark energy density $\Omega_{\rm de} \to 0$ remains the dominant factor. More importantly, the constraints at low redshift ($z < 1$) remain tight and centered on $w = -1$. When adopting the DESI prior for $\Omega_{m,0}$, we find a stronger constrain than that derived using the full sample without excluding the noisiest measurements (Equation~\eqref{eq:w0real}):
\begin{equation}
    w(0) = -0.988 \pm 0.101,
\end{equation}
reinforcing the conclusion that the local universe is consistent with a cosmological constant. This test demonstrates that while the inclusion of all available data is statistically complete, the future of model-independent reconstructions lies in the acquisition of high-precision CC measurements rather than simply increasing the data sample.
\begin{figure}[h]
\centering
\includegraphics[width=12cm]{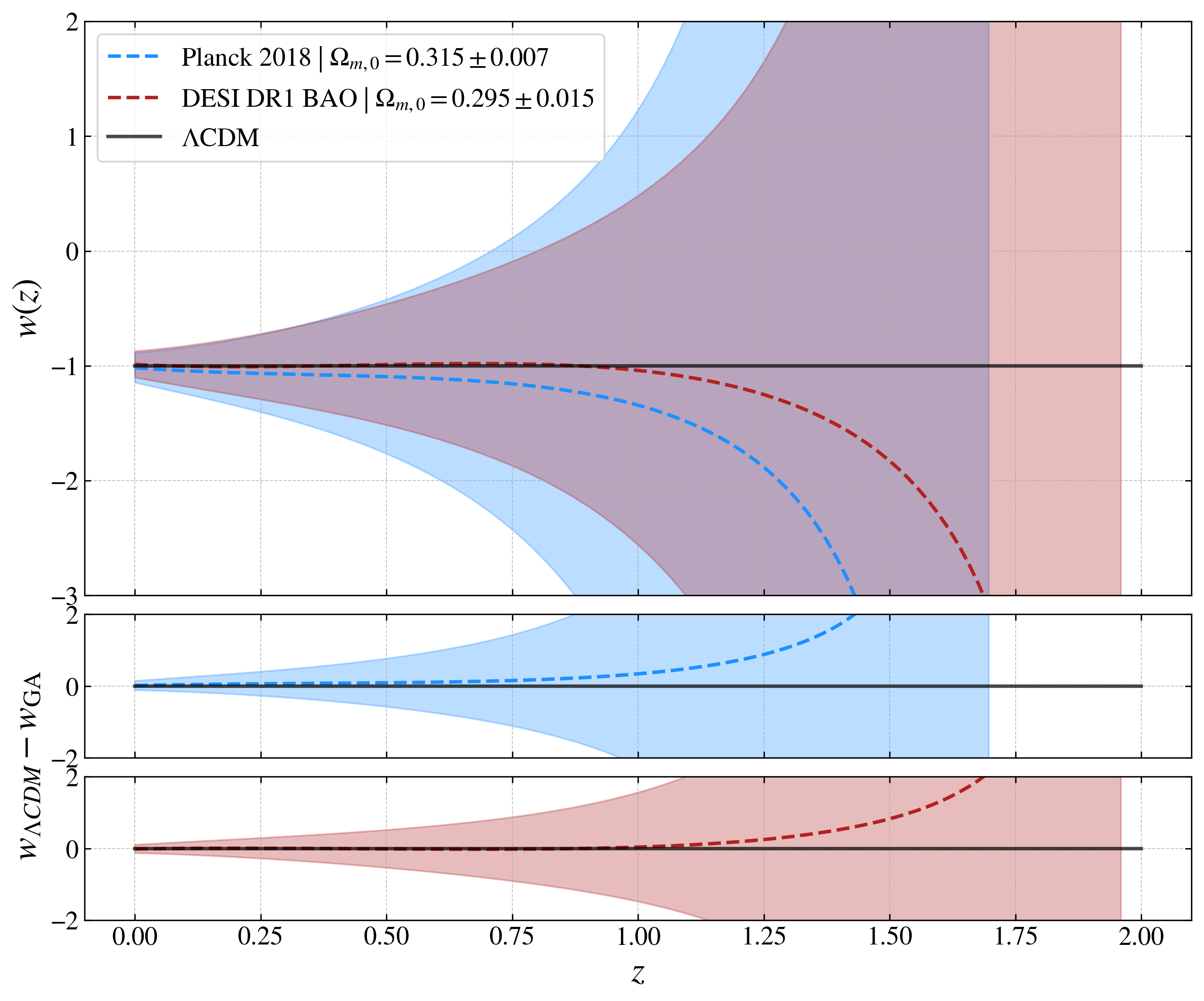}
\caption{Reconstruction of the dark energy equation of state $w(z)$ obtained from the quality-cut CC subset. As in Figure \ref{fig:real_w}, the blue dashed curve corresponds to imposing the Planck 2018 prior on $\Omega_{m,0}$, while the red dashed curve adopts the DESI DR1 BAO prior. Removing the noisiest measurements leads to a slightly more stable behavior at intermediate redshifts, while retaining the tight constraints at $z < 1$, which continue to be centered on the cosmological constant value $w = -1$ (solid black line). At high redshift ($z > 1.5$), the confidence regions still broaden because the dark energy density tends to zero ($\Omega_{\rm de} \to 0$), reinforcing that the weakening of constraints in this regime is inherent to the chosen parametrisation rather than driven by low-quality data.}
\label{fig:less_w}
\end{figure}

\section{Hyperparameter configurations setup}\label{app:ga_hyperparams}
Table~\ref{table:ga_hyperparams} summarises the GA hyperparameters used for the two reconstructions discussed in this work: the validation test function $f(x)$ (Section~\ref{ssec:f_mock}) and the CC reconstruction of the Hubble rate $H(z)$ (Section~\ref{ssec:H_real} and Section~\ref{ssec:H_mock}). For each target quantity we run the GA over an ensemble of $N_{\rm conf}$ configurations, which differ by the adopted grammar choice and by stochastic realisations (random seeds). The GA population size ($N_{\rm pop}$), the maximum number of generations ($N_{\rm gen}$), and the genetic operators rates (selection, crossover, mutation) are fixed within each reconstruction setup.
\begin{table}[!htbp]
\centering
\small
\setlength{\tabcolsep}{4pt}
\renewcommand{\arraystretch}{1.12}
\begin{tabularx}{\textwidth}{@{}>{\raggedright\arraybackslash}p{0.34\textwidth}
                        >{\centering\arraybackslash}X
                        >{\centering\arraybackslash}X
                        >{\centering\arraybackslash}X@{}}
\toprule
Hyperparameter & $f(x)$ & $H(z)$ & $\mu(z)$ \\
\midrule
Maximum generations $N_{\rm gen}$ & 3000 & 3000 & 500 \\
Population size $N_{\rm pop}$     & 100  & 100  & 40  \\
\addlinespace

Grammar set &
{\ttfamily [poly],\newline \ttfamily [cpl],\newline \ttfamily [poly,cpl]} &
{\ttfamily [poly,cpl,exp] + perm.\newline \ttfamily [poly,exp] + perm.\newline \ttfamily [poly], [exp], [cpl]} &
{\ttfamily [log,cpl]} \\
\addlinespace

Selection rate  & 0.30 & 0.30 & 0.40 \\
Crossover rate  & 0.85 & 0.85 & 0.95 \\
Mutation rate   & 0.85 & 0.85 & 0.95 \\
Number of random seeds $N_{\rm seeds}$ & 33  & 10  & 80 \\
Number of configurations $N_{\rm conf}$ & 99 & 150 & 80 \\
\bottomrule
\end{tabularx}
\caption{Hyperparameter setup adopted for the GA reconstructions. Here ``perm.'' indicates that all orderings (permutations) of the listed building blocks are included.}
\label{table:ga_hyperparams}
\end{table}
The grammar sets define the fundamental mathematical building blocks available to the algorithm for constructing candidate functions. In Table~\ref{table:ga_hyperparams}, these sets are denoted by their abbreviations:
\begin{itemize}
    \item \texttt{poly} represents standard polynomial terms (e.g., $x$, $x^2$, $x^3$),
    \item \texttt{exp} represents exponential functions (e.g., $e^x$, $e^{-x}$),
    \item \texttt{cpl} incorporates terms proportional to $x/(1+x)$, inspired by the Chevallier-Polarski-Linder parametrisation (\cite{Chevallier:2001, Linder:2003}),
    \item \texttt{log} represents logarithm functions (e.g., $\log{x}$).
\end{itemize}
The algorithm generates candidate functions by combining these building blocks using standard arithmetic operators.
The specific hyperparameters values, particularly the relatively high crossover and mutation probabilities, were adopted based on previous empirical tuning. Although standard GA implementations typically employ lower mutation rates, we found that high mutation probabilities help maintain genetic diversity in the population, forcing continued exploration of the functional space and reducing the risk of premature convergence to local minima.

\end{document}